\documentclass[nolinenumbers,trackchanges]{aastex701}

\begin{document}

\title{Dust Without Na I D Trace: The Case of Highly Attenuated Type Ib SN 2024vjc}

\author[orcid=0009-0000-2457-279X]{Javier Silva-Farf\'an}
\affiliation{Department of Astronomy, Kyoto University, Kitashirakawa-Oiwake-cho, Sakyo-ku, Kyoto, 606-8502. Japan}
\email[show]{javier@kusastro.kyoto-u.ac.jp}

\author[orcid=0000-0003-2611-7269]{Keiichi Maeda}
\affiliation{Department of Astronomy, Kyoto University, Kitashirakawa-Oiwake-cho, Sakyo-ku, Kyoto, 606-8502. Japan}
\email[]{keiichi.maeda@kusastro.kyoto-u.ac.jp}

\author[orcid=0000-0002-7259-4624]{Andrea Pastorello}
\affiliation{INAF - Osservatorio Astronomico di Padova, Vicolo dell'Osservatorio 5, I-35122, Padova, Italy}
\email[]{andrea.pastorello@inaf.it}

\author[orcid=0000-0002-1132-1366]{Hanindyo Kuncarayakti}
\affiliation{Tuorla Observatory, Department of Physics and Astronomy, FI-20014 University of Turku, Finland}
\affiliation{Finnish Centre for Astronomy with ESO (FINCA), FI-20014 University of Turku, Finland}
\email[]{kuncarayakti@gmail.com}  

\author[orcid=0000-0002-3933-7861]{Takashi Nagao}
\affiliation{Observatory of Japan, National Institutes of Natural Sciences, 2-21-1 Osawa, Mitaka, Tokyo, 181-8588, Japan}
\affiliation{Tuorla Observatory, Department of Physics and Astronomy, FI-20014 University of Turku, Finland}
\email[]{tnagao90@gmail.com}  

\author[orcid=0000-0002-3334-4585]{Giorgio Valerin}
\affiliation{INAF - Osservatorio Astronomico di Padova, Vicolo dell'Osservatorio 5, I-35122, Padova, Italy}
\email[]{giorgio.valerin@inaf.it}  

\author[0000-0003-4254-2724]{Andrea Reguitti}
\email[]{andrea.reguitti@inaf.it}
\affiliation{INAF - Osservatorio Astronomico di Padova, Vicolo dell'Osservatorio 5, I-35122, Padova, Italy}
\affiliation{INAF - Osservatorio Astronomico di Brera, Via E. Bianchi 46, I-23807, Merate (LC), Italy}

\author[orcid=0000-0001-9942-277X]{Dino Pierluigi Fugazza}
\affiliation{INAF - Osservatorio Astronomico di Brera, Via E. Bianchi 46, I-23807, Merate (LC), Italy}
\email[]{dino.fugazza@inaf.it}  

\author[0000-0003-0006-0188]{Giuliano Pignata}
\email[]{pignago@gmail.com}
\affiliation{Instituto de Alta Investigaci\'on, Universidad de Tarapac\'a, Casilla 7D, Arica, Chile}

\author[0000-0003-1546-6615]{Jesper Sollerman}
\email[]{jesper@astro.su.se}
\affiliation{The Oskar Klein Centre, Department of Astronomy, Stockholm University, SE-106 91 Stockholm , Sweden}

\author[0000-0002-5221-7557]{C. Ashall}
\email[]{cashall@hawaii.edu}
\affiliation{Institute for Astronomy, University of Hawai'i at Manoa, 2680 Woodlawn Dr., Hawai'i, HI 96822, USA}

\author[0000-0003-3939-7167]{ Tom\'as E. Müller-Bravo}
\email[]{t.e.muller-bravo@tcd.ie}
\affiliation{School of Physics, Trinity College Dublin, The University of Dublin, Dublin 2, Ireland}
\affiliation{Instituto de Ciencias Exactas y Naturales (ICEN), Universidad Arturo Prat, Chile}

\author[0000-0003-0227-3451]{Joseph P. Anderson}
\email[]{janderso@eso.org}
\affiliation{European Southern Observatory, Alonso de Córdova 3107, Vitacura, Casilla 19001, Santiago, Chile}

\author[0000-0003-2375-2064]{Claudia P. Guti\'errez}
\email[]{cgutierrez@ice.csic.es}
\affiliation{Institute of Space Sciences (ICE, CSIC), Campus UAB, Carrer de Can Magrans, s/n, E-08193 Barcelona, Spain}

\author[0000-0002-8041-8559]{Priscila J. Pessi}
\email[]{pjpessi@gmail.com}
\affiliation{Astrophysics Division, National Centre for Nuclear Research, Pasteura 7, 02-093 Warsaw, Poland}

\author[0000-0001-8257-3512]{Erkki Kankare}
\email[]{erkki.kankare@utu.fi}
\affiliation{Tuorla Observatory, Department of Physics and Astronomy, FI-20014 University of Turku, Finland}

\author[0009-0006-0165-6986]{Niko Pyykkinen}
\email[]{njpyyk@utu.fi}
\affiliation{Tuorla Observatory, Department of Physics and Astronomy, FI-20014 University of Turku, Finland}
\affiliation{Nordic Optical Telescope, Rambla José Ana Fernández, Pérez 7, E38711 Breña Baja, Spain}

\author[0000-0002-1650-1518]{Mariusz Gromadzki}
\email[]{mariusz.gromadzki@gmail.com}
\affiliation{Astronomical Observatory, University of Warsaw, Al. Ujazdowskie 4, 00-478 Warszawa, Poland}

\author[0000-0002-1066-6098]{Ting-Wan Chen}
\email[]{twchen@gm.astro.ncu.edu.tw}
\affiliation{Graduate Institute of Astronomy, National Central University, 300 Jhongda Road, 32001 Jhongli, Taiwan}

\author[0000-0002-1022-6463]{Thomas M. Reynolds}
\email[]{treynolds1729@gmail.com}
\affiliation{Tuorla Observatory, Department of Physics and Astronomy, FI-20014 University of Turku, Finland}
\affiliation{Niels Bohr Institute, University of Copenhagen, Jagtvej 128, DK-2200, Copenhagen N, Denmark}
\affiliation{Cosmic Dawn Center (DAWN)}

\author[0000-0003-1450-0869]{Irene Salmaso}
\email[]{irene.salmaso@inaf.it}
\affiliation{INAF–Osservatorio Astronomico di Capodimonte, Salita Moiariello
16, 80131 Napoli, Italy}
\affiliation{INAF - Osservatorio Astronomico di Padova, Vicolo dell'Osservatorio 5, I-35122, Padova, Italy}

\author[0000-0002-3653-5598]{Avishay Gal-Yam}
\email[]{avishay.gal-yam@weizmann.ac.il}
\affiliation{Department of Particle Physics and Astrophysics, Weizmann Institute of Science, 234 Herzl St, 7610001 Rehovot, Israel}



\accepted{2026 July 23}
\submitjournal{The Astrophysical Journal}

\begin{abstract}
Understanding dust attenuation toward extragalactic transients is critical for recovering their intrinsic properties and probing the local environments of distant galaxies. A popular diagnostic is the Na~{\sc i}~D absorption equivalent width widely applied to extragalactic transients. In this paper, we present early-time optical and near-infrared observations of the Type Ib supernova (SN) SN 2024vjc, followed for $\sim$130 days post-explosion. SN~2024vjc exhibits only weak Na~I~D absorption, indicating only a modest host attenuation with $E (B-V)_{\rm host} \sim 0.18$ mag; if this argument were applied, SN~2024vjc would be a peculiar, faint, and red SN Ib while showing the light-curve shape and spectral evolution broadly consistent with those of canonical SNe Ib. We show that this is not the case; SN-based diagnostics (intrinsic color templates, color-curve evolution, and spectral dereddening) together with the Balmer decrement indicate substantial attenuation, $E(B-V) \sim 0.45$ $-$ $0.8$\,mag. The weak Na~{\sc i}~D absorption may result from photoionization of neutral sodium by the intense radiation field of a young \ion{H}{2} region at the explosion site; this scenario is directly supported by the detection of narrow H$\alpha$, [\ion{N}{2}], and H$\beta$ emission lines and a blue continuum excess in late-time spectroscopy. SN~2024vjc represents a clear counterexample to the commonly assumption that weak or absent Na~I~D absorption implies negligible host-galaxy attenuation, and highlights the importance of employing multiple, independent attenuation diagnostics.

\end{abstract}



\section{Introduction} 
Supernovae (SNe) are the terminal explosions of stars, reaching luminosities comparable to those of their host galaxies. They are broadly classified into thermonuclear supernovae (SNe Ia) of white dwarfs and core-collapse supernovae (CCSNe) of massive stars ($M_{\rm ZAMS} \geq 8$ $M_\odot$). 
Among CCSNe, an important subclass is formed by the stripped-envelope supernovae (SESNe), whose progenitors have lost part or most of their outer hydrogen and/or helium layers prior to explosion. 
The degree of stripping determines the SESN subtype. SNe IIb retain a small amount of hydrogen, visible in early-time spectra but fading over time. SNe Ib have completely lost their hydrogen envelopes while retaining helium, producing spectra with prominent helium lines. SNe Ic have been stripped of both hydrogen and helium, exhibiting neither feature in their spectra \citep[see][]{1995ApJ...450L..11F, 1997ApJ...491..375C, 2001AJ....121.1648M, 2017hsn..book..195G, 2019NatAs...3..717M}.

The stripping of SESNe progenitors is believed to occur through two main evolutionary channels: (i) single massive stars, such as Wolf–Rayet stars, that remove their outer layers through strong stellar winds \citep{1986ApJ...302L..59B, 1995ApJ...448..315W}, or (ii) binary systems, where a companion star removes the progenitor’s envelope via mass transfer \citep{1992ApJ...391..246P}. 
Observations suggest that the major pathway to `classical' SESNe (Types IIb/Ib/Ic) is likely the binary channel \citep[e.g.,][]{lyman2016MNRAS.457..328L,fang2019NatAs...3..434F}, but multiple evolutionary channels probably contribute to observed diversities in the photometric and spectroscopic properties beyond the classical categories, as indicated by recent discoveries of various flavors of `hydrogen-free' interacting SNe. 
These include Type~Ibn SNe, which interact with helium-rich circumstellar material (CSM) \citep{pastorello2007Natur.447..829P}; Type~Icn SNe, characterized by interaction with carbon/oxygen-rich shells \citep{gal-yam2022Natur.601..201G}; and the recently proposed Type~Ien SNe (e.g., SN~2021yfj), which reveal the most interior stellar layers rich in oxygen, silicon, and sulfur \citep{schulze2025Natur.644..634S}. While these examples are characterized by a `confined' hydrogen-poor CSM with a typical scale of $\sim 10^{15}$ cm \citep{maeda2022ApJ...927...25M}, another class of `SNe Ic-CSM' show signatures of a dense and extended hydrogen-poor CSM \citep{kuncarayakti2022ApJ...941L..32K, kuncarayakti2023A&A...678A.209K,maeda2026PASJ...78L...1M}.


Studying SNe provides insight into a wide range of astrophysical topics, including high-energy processes, the final stages of stellar evolution, the interstellar medium (ISM) of the  host-galaxy, the metallicity of the local environment, and the measurement of cosmic distances. To obtain reliable photometric measurements that can support these investigations, it is important to know the intrinsic luminosity and spectral energy distribution (SED) of the SNe being studied. Therefore, understanding the properties of the Milky Way (MW) attenuation and the `host attenuation' (that could involve both the local and galactic-scale environment) is of fundamental importance. 

The standard prescription for the Galactic attenuation is well established, using a combination of dust maps \citep[such as][]{1998ApJ...500..525S, 2011ApJ...737..103S} and several reddening laws ranging from the foundational \citet{1989ApJ...345..245C} (CCM89) parameterization to the widely used \citet{1999PASP..111...63F} (F99) law. On the other hand, correcting for the host attenuation presents a greater challenge -- the properties of dust grains that produce the extinction are not necessarily the same as those found in the MW environment \citep[e.g.,][]{kawabata2014ApJ...795L...4K,nagao2022ApJ...941L...4N}. 

Methods to estimate the host attenuation based on observations of extragalactic SNe (or transients, in general) are largely divided into two categories; one relying on knowledge of the intrinsic luminosity or color of the SN to be studied, and the other using an independent indicator without assuming the SN properties. The former, the SN-based method include intrinsic color comparisons \citep{2011ApJ...741...97D, 2015A&A...574A..60T,2018A&A...609A.135S} to obtain the $E(B-V)$$_{host}$ value. One of the most widely used techniques for the SN-independent method to estimate $E(B-V)$$_{host}$ is from the equivalent width (EW) of the narrow Na I D absorption lines \citep[e.g.,][]{2012MNRAS.426.1465P}, despite several limitations; the relation between EW$_{\mathrm{NaID}}$ and $E(B-V)$ exhibits a large scatter for EW$_{\mathrm{NaID}}$ values of $\geq$ 1 \AA, and it becomes inaccurate when the spectra are of low resolution \citep{2011MNRAS.415L..81P}. 

These methods have been used for SNe Ia, making use of their standardized nature in intrinsic luminosity and color \citep{phillips1999AJ....118.1766P}. \citet{2013ApJ...779...38P} showed that Na\,\textsc{i}\,D is not a reliable indicator of host-galaxy attenuation for SNe~Ia, concluding that the absence of Na\,\textsc{i}\,D absorption in high signal-to-noise spectra is \textit{consistent with} low dust attenuation [i.e.\ $E(B-V)_{\rm host} \approx 0$ mag]. In addition, the attenuation within the SN Ia host galaxies (or the local SN environment) has been inferred to be different from the MW law (as represented by CCM89), typically having a low value of $R_{V}$ $\equiv$ $A_{V}$/$E(B-V)$ \citep{2013ApJ...779...38P}. 

The situation is even more challenging for CCSNe; the progenitor environments, ionization conditions, and dust properties may differ substantially from those with which the Na I D -- $E(B–V)$ relation was originally calibrated. The intrinsic diversity in SESN photometric and spectroscopic properties does not allow accurate determination of $E(B-V)$ \citep{2023ApJ...955...71R}. Conversely, the uncertainty in the attenuation limits reliable estimation of the SN intrinsic properties such as the luminosity and $^{56}$Ni mass. Therefore, calibrating the Na I D -- $E(B–V)$ relation is important for understanding both the intrinsic properties of CCSNe and their environmental characteristics. In the present work, we test whether the assumption that weak or absent Na~{\sc i}~D absorption implies negligible host-galaxy attenuation holds in the context of SESNe. A single well-documented counterexample is sufficient to demonstrate that this diagnostic cannot be universally relied upon.

In this work, we study photometric and spectral properties of the type Ib SN 2024vjc, focusing on its attenuation. While we find various lines of evidence for it being a standard SN Ib suffering from significant attenuation, Na I D absorption is barely detected in our best signal-to-noise spectrum and absent in the other spectra. SN 2024vjc thus demonstrates that the absence of detectable Na I D absorption cannot be taken as definitive evidence that an SN is unaffected by  host-galaxy  attenuation, highlighting the importance of developing and applying methods to estimate host-galaxy attenuation that do not rely on measurements of EW$_{\mathrm{NaID}}$. 
In Section \ref{sec:observations}, we describe the observational data used in this work. Section \ref{sec:light_curve} presents the optical and near-infrared (NIR) photometric properties of SN 2024vjc. In Section \ref{sec:Spectra}, we analyze the spectroscopic evolution of the SN and present \textsc{SYN++} spectral modeling. Section \ref{sec:attenuation} is dedicated to the analysis of host-galaxy attenuation using multiple independent diagnostics. In Section \ref{sec:discussion}, we discuss the implications of our results and the possible physical origin of the strong attenuation inferred for SN 2024vjc. Finally, our conclusions are presented in Section~\ref{sec:conclusion}.

\section{Observations} \label{sec:observations}

\subsection{SN 2024\lowercase{vjc} discovery and  host-galaxy }

SN 2024vjc was discovered on 2024 September 13 (MJD=60\,565.2) by the Asteroid Terrestrial-impact Last Alert System \citep[ATLAS, ][]{2024TNSTR3476....1T} El Sauce telescope. An initial spectroscopic classification suggested SN IIb classification, which was subsequently revised to SN Ib on 2024 September 14 using the ESO Faint Object Spectrograph and Camera v.2 (EFOSC2; \citealt{1984Msngr..38....9B, 2008Msngr.132...18S, 2024TNSCR3507....1A}) mounted on the 3.58\,m New Technology Telescope (NTT) at La Silla Observatory. We estimate the explosion epoch by fitting a fireball model, $f(t) \propto (t - t_{\mathrm{exp}})^{2}$, to the early ATLAS $o$-band forced photometry in flux space. Over a 4-day window from the first detection, we obtain $t_{\mathrm{exp}} = 60\,559.5 \pm 1.5$ MJD, where the uncertainty combines the statistical error from bootstrap resampling with a systematic from varying the fit window between 3 and 6 days.

\begin{figure}[ht!]
\centering
\includegraphics[width=0.6\linewidth]{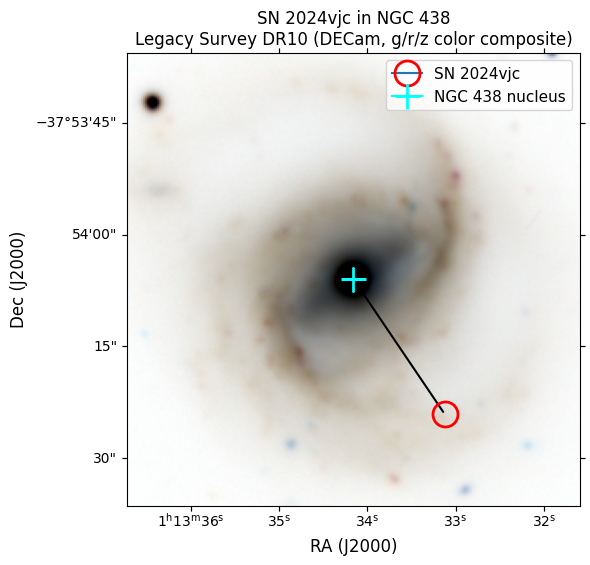}
\caption{Legacy Survey DR10 (DECam) g/r/z color composite image of NGC 438, host galaxy of SN 2024vjc. The nucleus of NGC 438 is marked by a cyan cross, while the location of SN 2024vjc is indicated by a red circle.}
\label{fig:NGC438}
\end{figure}

SN 2024vjc is located at $\alpha = 01^{\mathrm h}13^{\mathrm m}33.123^{\mathrm s}$, $\delta = -37^{\circ}54^{\prime}24.08^{\prime\prime}$ (J2000), within a spiral arm of the  host-galaxy  NGC 438, at a projected galactocentric distance of $\sim$21.9$^{\prime\prime}$ from the galaxy nucleus (Figure \ref{fig:NGC438}).  We adopt a heliocentric redshift of $z = 0.01156 \pm 0.00005$ for NGC~438, as listed in the SIMBAD database\footnote{\url{https://simbad.u-strasbg.fr/simbad/}}. Assuming a flat $\Lambda$CDM cosmology with $H_0 = 70$~km~s$^{-1}$~Mpc$^{-1}$ and $\Omega_m = 0.3$, this corresponds to a luminosity distance of $\sim 49.7$~Mpc. The Galactic attenuation toward SN 2024vjc is estimated using the dust maps of \citet{2011ApJ...737..103S}, adopting the \citet{1989ApJ...345..245C} attenuation law with $R_V^{\rm MW} = 3.1$, which yields $A_V^{\rm MW} = 0.037$\,mag.

\subsection{ATLAS photometry}


We obtained ATLAS photometry of SN~2024vjc in cyan and orange bands by querying the ATLAS Forced Photometry server \citep{2018PASP..130f4505T, 2020PASP..132h5002S}, covering the period from 2024-07-14 to 2025-05-22 (approximately 60 days before explosion to 250 days after explosion). The forced-photometry light curve was then stacked on a nightly basis using the script provided by \citet{Young_plot_atlas_fp}, to improve the signal-to-noise ratio and reduce the scatter of the measurements.

\subsection{REM photometry}

Follow-up observations of SN~2024vjc were carried out with the Rapid Eye Mount (REM) telescope \citep{2003Msngr.113...40C} at La Silla Observatory. Data were obtained using both available instruments: REMIR, which operates in the infrared and provides imaging in the $J$, $H$, and $K'$ bands, and ROSS2, which operates in the optical and provides imaging in the Sloan $g'$, $r'$, $i'$, and $z'$ bands. Observations started on 2024-09-20 (approximately 10 days after $t_{exp}$) and continued until 2024-11-18 (approximately 70 days after $t_{exp}$).

Astrometric and photometric calibrations, as well as the SN magnitude measurements, were performed using routines from the \textsc{ecsnoopy} package \citep{ecsnoopy}\footnote{\textsc{ecsnoopy} is a Python package developed by E.~Cappellaro for supernova photometry, based on point-spread function fitting and template subtraction; a detailed description is available at \url{http://sngroup.oapd.inaf.it/ecsnoopy.html}.}. Instrumental magnitudes were calibrated using photometric zero points derived from local reference stars at each epoch. Optical calibrations were performed using photometry from the AAVSO Photometric All-Sky Survey \citep[APASS;][]{apass}, while NIR calibrations were based on the 2MASS catalog \citep{2003yCat.2246....0C}.

\subsection{SPECULOOS, EFOSC2, and VST photometry}

Additional optical follow-up observations were obtained with the SPECULOOS telescopes \citep{2018SPIE10700E..1ID}, EFOSC2 mounted on the ESO New Technology Telescope, and OmegaCAM mounted on the VLT Survey Telescope (VST) \citep{2011Msngr.146....8K}. The VST observations were obtained in the $uBVgriz$ filters, although the final epoch only includes observations in the $uBV$ bands.

The SPECULOOS and VST data were reduced using template subtraction techniques in order to minimize host-galaxy contamination. Deep archival images from the Dark Energy Survey \citep[DES;][]{2018ApJS..239...18A} were used as templates for the Sloan $griz$ bands, while archival images from SkyMapper \citep{2024PASA...41...61O} were used for the $V$-band template subtraction. For the $u$ band, archival VST images were employed as templates. PSF photometry was then performed on the difference images using the same technique adopted for the REM data 

Photometric calibration of the subtracted images was performed using local field stars calibrated against the DES and SkyMapper catalogs. The resulting light curves show good agreement with the REM photometry after subtraction of the host-galaxy contribution.

\subsection{Spectra}



We obtained the majority of the optical spectroscopic observations with EFOSC2 as part of the extended Public ESO Spectroscopic Survey of Transient Objects and their Environmentn (ePESSTO+; \citealt{2015A&A...579A..40S, 2019eeu..confE..15C}). The EFOSC2 spectra were reduced using the dedicated \textsc{pessto} pipeline (Section~4 of \citealt{2015A&A...579A..40S}), a Python-based framework designed for uniform processing of survey data. The pipeline performs standard reduction steps including bias subtraction, flat-fielding, and the removal of cosmic rays using a Python implementation of the \textsc{l.a.cosmic} algorithm \citep{2001PASP..113.1420V}. Wavelength calibration is performed using He+Ar arc lamp exposures. Flux calibration is achieved with spectrophotometric standard stars observed multiple times per night with the same instrument configuration. 

We acquired spectra with the Gemini Multi-Object Spectrograph (GMOS) on the 8 m Gemini South telescope on 2024-09-15, 2024-09-21, and 2025-01-29. For the first two epochs, we used 1.0''-width slit with the B480 grism with the central wavelength set at 5400 \AA, and a total of 1200 s exposure was divided into 4 sequences with spectral dithering to correct for the detector gaps. For the late-time spectrum taken with Gemini-S/GMOS on 2025-01-29, we used the same setup but with a total exposure of 1800 s. 
The data from Gemini/GMOS were reduced using the Data Reduction for Astronomy from Gemini Observatory North and South (DRAGONS) software \citep{2023RNAAS...7..214L} using standard procedures of bias subtraction, flat-fielding, wavelength and flux calibration, resulting in reduced spectra with wavelength range of $\sim380-740$ nm.

Three additional spectra were obtained with the Andalucia Faint Object Spectrograph and Camera (ALFOSC) mounted on the 2.56\,m Nordic Optical Telescope (NOT) at Roque de los Muchachos Observatory, La Palma, under the NOT Un-biased Transient Survey 2 (NUTS2) program. The ALFOSC spectra were reduced using the \texttt{ALFOSCGUI}\footnote{\url{http://sngroup.oapd.inaf.it/foscgui.html}} pipeline, which carries out bias subtraction, flat-fielding, wavelength calibration with arc lamps, and flux calibration using a spectrophotometric standard star observed on the same night.

Table~\ref{tab:spectra} lists the spectroscopic observations of 
SN~2024vjc.

\begin{deluxetable*}{ccccccccc}
\tablecaption{Spectroscopic Observations of SN 2024vjc \label{tab:spectra}}
\tablehead{
\colhead{Obs. Date} & \colhead{MJD} & \colhead{Phase [days]} & \colhead{Telescope} & \colhead{Instrument} & \colhead{Grism} &  \colhead{Exp. Time [s]} & \colhead{FWHM$_{\rm INST}$ [\AA]} & \colhead{Airmass}}
\startdata
2024-09-14 & 60\,567 & $-14$ & ESO-NTT      & EFOSC2 & Gr\#13 & 1800 & 23.0  & 1.07 \\
2024-09-15 & 60\,568 & $-13$ & Gemini-South & GMOS-S & B480   & 1200 & 6.25  & 1.01 \\
2024-09-21 & 60\,574 & $-7$  & Gemini-South & GMOS-S & B480   & 1200 & 6.25  & 1.38 \\
2024-09-27 & 60\,580 & $-1$ & NOT          & ALFOSC & Gr\#4  & 1800 & 14.3  & 2.51 \\
2024-10-03 & 60\,586 & $+5$  & ESO-NTT      & EFOSC2 & Gr\#13 & 1800 & 34.45 & 1.21 \\
2024-10-09 & 60\,592 & $+11$ & NOT          & ALFOSC & Gr\#4  & 2700 & 14.3  & 2.58 \\
2024-10-19 & 60\,602 & $+21$ & NOT          & ALFOSC & Gr\#4  & 3600 & 14.3  & 2.54 \\
2024-11-06 & 60\,620 & $+39$ & ESO-NTT      & EFOSC2 & Gr\#13 & 2700 & 23.0  & 1.06 \\
2024-11-14 & 60\,628 & $+47$ & ESO-NTT      & EFOSC2 & Gr\#13 & 2100 & 23.0  & 1.11 \\
2024-12-01 & 60\,645 & $+64$ & ESO-NTT      & EFOSC2 & Gr\#13 & 2700 & 23.0  & 1.07 \\
2024-12-22 & 60\,666 & $+85$ & ESO-NTT      & EFOSC2 & Gr\#13 & 2700 & 23.0  & 1.68 \\
2025-01-08 & 60\,683 & $+102$ & ESO-NTT     & EFOSC2 & Gr\#13 & 2700 & 23.0  & 1.39 \\
2025-01-29 & 60\,704 & $+123$ & Gemini-South & GMOS-S & B480  & 1800 & 6.25  & 1.71 \\
\enddata
\tablecomments{Phases are relative to $g$-band maximum. Instrumental resolution (FWHM$_{\rm inst}$) is estimated from the slit width and nominal dispersion.}
\end{deluxetable*}

\section{Light curve} \label{sec:light_curve}

\begin{figure}[ht!]
\centering
\includegraphics[width=0.8\linewidth]{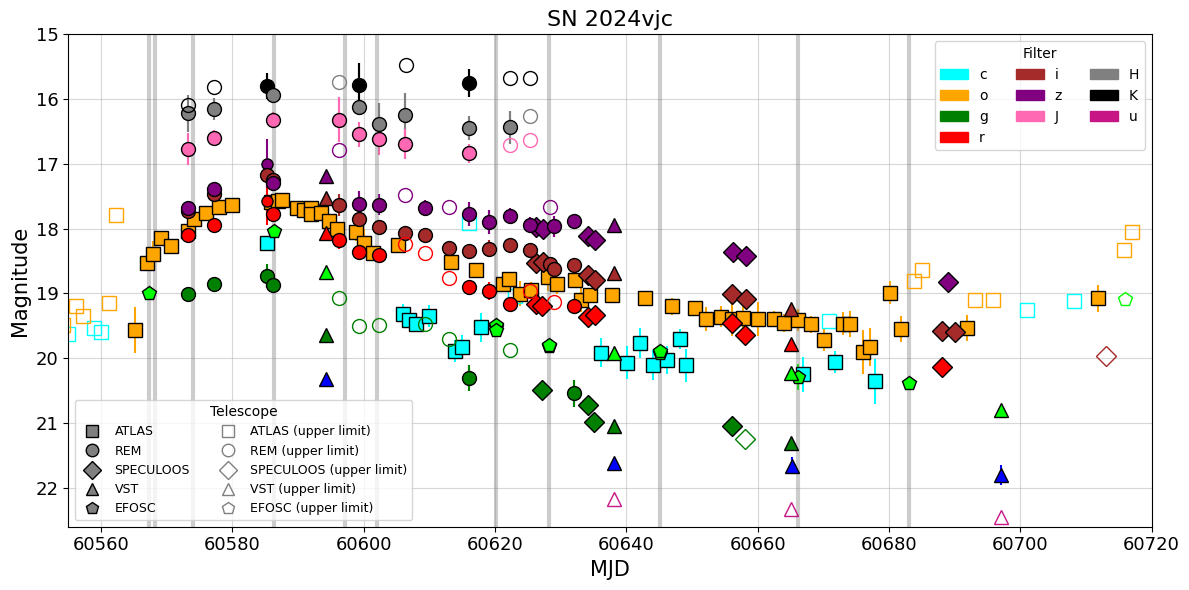}
\caption{Multi-band optical and NIR light curves of the Type~Ib supernova SN~2024vjc corrected by MW attenuation. Photometry is shown for the $u, B, V, g, r, i, z, o, c, J, H, K$-bands, color-coded by band as indicated in the legend. Marker shapes represent different facilities: ATLAS (squares), REM (circles), SPECULOOS (diamonds), VST/OmegaCAM (upward triangles), and EFOSC (pentagons). Open markers of the corresponding shape denote non-detections (upper limits) from the same Telescope. Magnitudes are corrected for MW extinction. Vertical gray bars indicate the epochs of spectroscopic observations.}
\label{fig:sn24vjc_lc}
\end{figure}

Using data from ATLAS forced photometry, the REM telescope, SPECULOOS, VST/OmegaCAM, and EFOSC2, we present the multiband light curve (LC) of SN 2024vjc in Figure \ref{fig:sn24vjc_lc}. All magnitudes are corrected for MW extinction. The dataset includes photometry in the $u, B, V, g, r, i, z, o, c, J, H, K$ bands, though the VST $u$-band yields only upper limits. Gray vertical lines correspond to the epochs of spectroscopic observations. 

The REM $g$-band shows only a few detections, with most measurements being non-detections. Overall, the redder bands are systematically brighter compared to the bluer bands from the pre-maximum phase through the radioactive decay tail. We note a small excess in the $i$- and $z$-band light curves around MJD $\sim$60\,625 ($\sim$65~d after explosion), accompanied by a corresponding peak in the $r-i$ color evolution, confirmed independently by both REM and VST photometry.

To estimate the peak epoch, maximum brightness, and the decline rate $\Delta m_{15}$ for each band, we follow the methodology of \citet{2014ApJS..213...19B}. We perform a second-degree polynomial fit to each monochromatic light curve (LC) using a Monte Carlo approach centered on the expected peak. We verified these polynomial fits against a smoothed cubic spline and found agreement within $1\sigma$ for the well-sampled $r$, $o$, and $i$ bands. Rise times are computed relative to the explosion epoch derived in Section~2.1.


\begin{deluxetable*}{lcccc}
\tablecaption{Light curve parameters of SN 2024vjc in different bands based on the method of \citet{2014ApJS..213...19B}. Rise times are computed relative to $t_{\mathrm{exp}} = 60559.5 \pm 1.5$ MJD. \label{tab:lc_bianco_params}}
\tablehead{
\colhead{Band} & \colhead{Peak Brightness [mag]} & \colhead{$t_{\mathrm{max}}$ [MJD]} & \colhead{$\Delta m_{15}$ [mag]} & \colhead{$t_{\mathrm{rise}}$ [days]}
}
\startdata
$g$ & 18.70$\pm$0.09 & 60\,581.1$\pm$1.0 & 1.22$\pm$0.24 & 21.6$\pm$1.8 \\
$r$ & 17.79$\pm$0.07 & 60\,584.3$\pm$1.1 & 0.61$\pm$0.19 & 24.8$\pm$1.8 \\
$i$ & 17.22$\pm$0.03 & 60\,585.4$\pm$0.6 & 0.76$\pm$0.09 & 25.9$\pm$1.6 \\
$z$ & 17.19$\pm$0.07 & 60\,587.2$\pm$4.6 & 0.56$\pm$0.19 & 27.7$\pm$4.8 \\
$o$ & 17.57$\pm$0.02 & 60\,585.0$\pm$0.5 & 0.68$\pm$0.09 & 25.5$\pm$1.6 \\
$J$$^{*}$ & 16.29$\pm$0.16 & 60\,589.1$\pm$14.0 & 0.45$\pm$0.25 & 29.6$\pm$14.1 \\
$H$$^{*}$ & 15.93$\pm$0.18 & 60\,587.1$\pm$18.2 & 0.38$\pm$0.23 & 27.6$\pm$18.2 \\
\enddata
\tablecomments{The $B$, $V$, $c$, and $K$ bands lack sufficient coverage around peak and are therefore omitted. $^{*}$NIR bands are loosely constrained by the available sampling; values should be interpreted with caution.}
\end{deluxetable*}

The derived parameters are summarized in Table~\ref{tab:lc_bianco_params}. We omit the $B$, $V$, $c$, and $K$ bands from this analysis due to insufficient coverage near the peak. We adopt the $g$-band peak as the reference epoch, MJD $g_{\mathrm{max}} = 60\,581.1$ (2024-09-28). As shown in Figure~\ref{fig:sn24vjc_lc}, the LCs in redder bands peak at progressively later times; this behavior is consistent with cooling ejecta \citep{2014ApJS..213...19B}. Furthermore, the peak brightness increases at longer wavelengths. This trend may be attributed to host-galaxy attenuation suppressing the blue flux or to the supernova possessing an intrinsically cool photosphere.

\begin{figure}[ht!]
\centering
\includegraphics[width=0.85\linewidth]{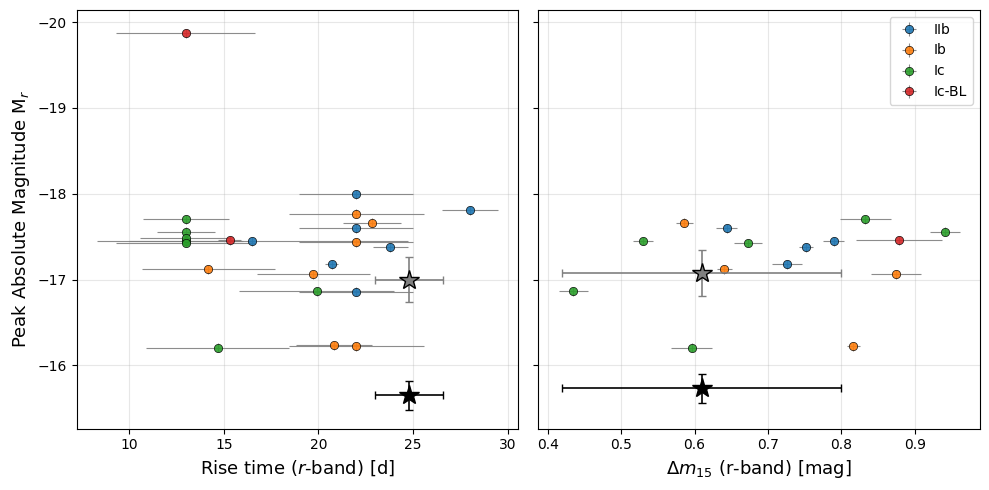}
\caption{\textit{Left:} Peak absolute $r$-band magnitude versus rise time. \textit{Right:} Peak absolute $r$-band magnitude versus decline rate parameter $\Delta m_{15}(r)$. The comparison sample comprises SESNe from the T18 sample, color-coded by subtype (IIb: blue, Ib: orange, Ic: green, Ic-BL: red). Galactic extinction has been corrected for using the \citet{2011ApJ...737..103S} dust maps with $R_V = 3.1$; no host-galaxy correction is applied to the comparison sample. SN~2024vjc is shown as a black star without host-galaxy attenuation correction, and as a gray star after applying a host attenuation correction of $E(B-V)_{\mathrm{host}} = 0.59$\,mag 
}
\label{fig:taddia_comparison}
\end{figure}

In Figure~\ref{fig:taddia_comparison}, we compare the $r$-band light-curve parameters of SN 2024vjc with those of the CSP-I SESN sample studied by \citet{2018A&A...609A.136T} (hereafter T18), which provides homogeneous multi-band photometry for a large sample of SESNe including SNe~IIb, Ib, Ic, and Ic-BL. For the T18 sample, we apply only Galactic extinction corrections using the \citet{2011ApJ...737..103S} dust maps with an assumed value of $R_V = 3.1$. This ensures our comparison does not depend on host-galaxy extinction assumptions. Rise times are computed as $t_{\mathrm{max}}(r) - t_{\mathrm{exp}}$, and $\Delta m_{15}(r)$ is estimated from a local polynomial fit to the $r$-band light curve between $+10$ and $+20$\,d relative to peak. The comparison includes the relation between the peak absolute magnitude, the decline rate ($\Delta m_{15}$), and the rise time. SN~2024vjc exhibits a typical decline rate but a rise time on the longer end of the SESN distribution, and appears as the faintest object in absolute magnitude when host-galaxy attenuation is not considered. If the host attenuation is corrected assuming $E(B-V)_{\mathrm{host}} = 0.59 \pm 0.09$\,mag (see Section~\ref{sec:attenuation}), SN~2024vjc falls into the luminosity range of the CSP-I SESN sample.

\begin{figure}[ht!]
\centering
\includegraphics[width=0.8\linewidth]{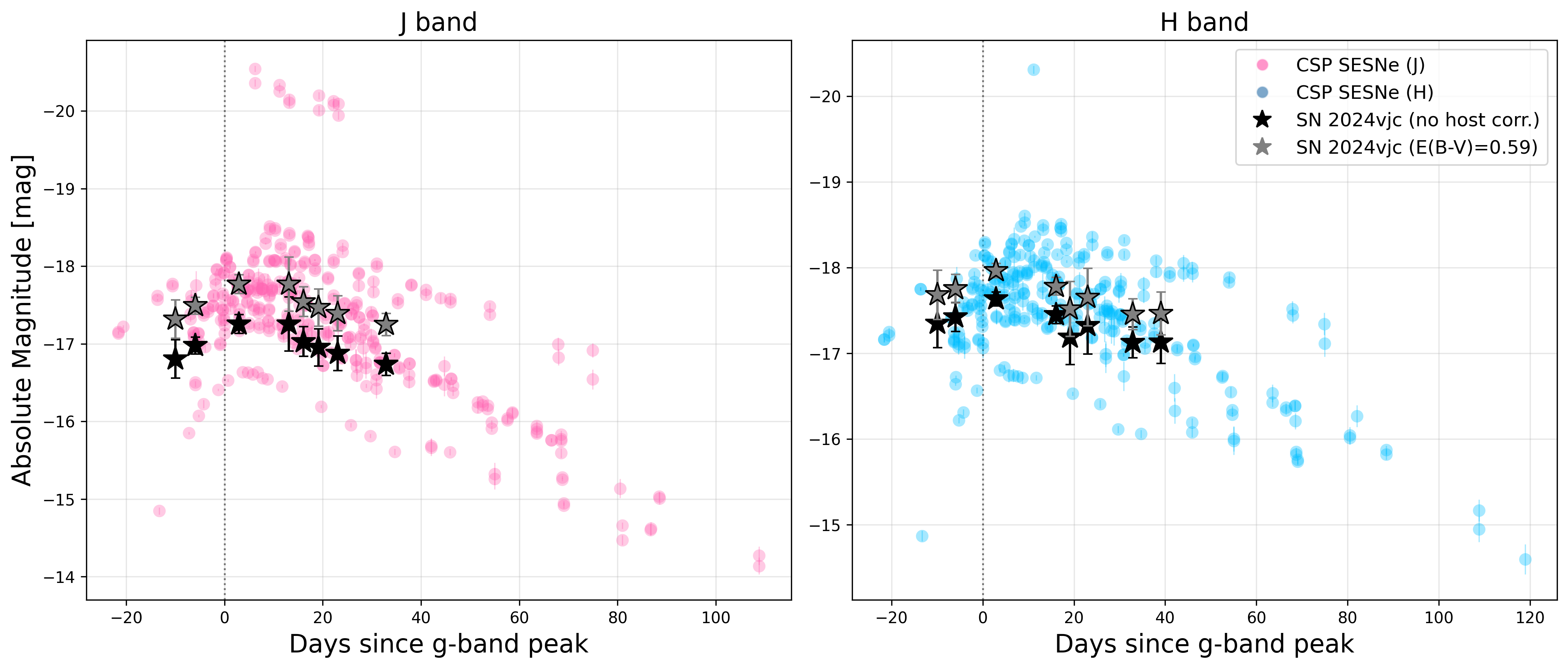}
\caption{Absolute NIR light curves of SN~2024vjc compared to the T18 sample. The left and right panels show the $J$ (pink points) and $H$ (blue points) bands, respectively, as a function of days since $g$-band maximum. Black and gray stars show SN~2024vjc without host-galaxy extinction correction and corrected for $E(B-V)_{host}$ = 0.59 mag respectively}
\label{fig:taddia_NIR_abs_mag}
\end{figure}

The NIR behavior of SN~2024vjc is further explored in Figure~\ref{fig:taddia_NIR_abs_mag}, where we compare the absolute $J$ and $H$-band light curves with the T18 sample. NIR bands are significantly less sensitive to dust extinction than their optical counterparts. Without accounting for host-galaxy attenuation, SN~2024vjc occupies the fainter end of the NIR distribution, consistent with our optical findings. However, in contrast to the optical properties, SN~2024vjc is not the faintest object in the NIR sample even without the host attenuation correction. After correcting for $E(B-V)_\mathrm{host}$ = $0.59$ mag, the NIR absolute magnitudes shift to $M_{J}\approx$ $-17.5$ and $M_{H}\approx$ $-17.8$ magnitudes near peak, placing SN~2024vjc well within the bulk of the CSP SESNe population in both bands.

\begin{figure}[ht!]
\centering
\includegraphics[width=0.7\linewidth]{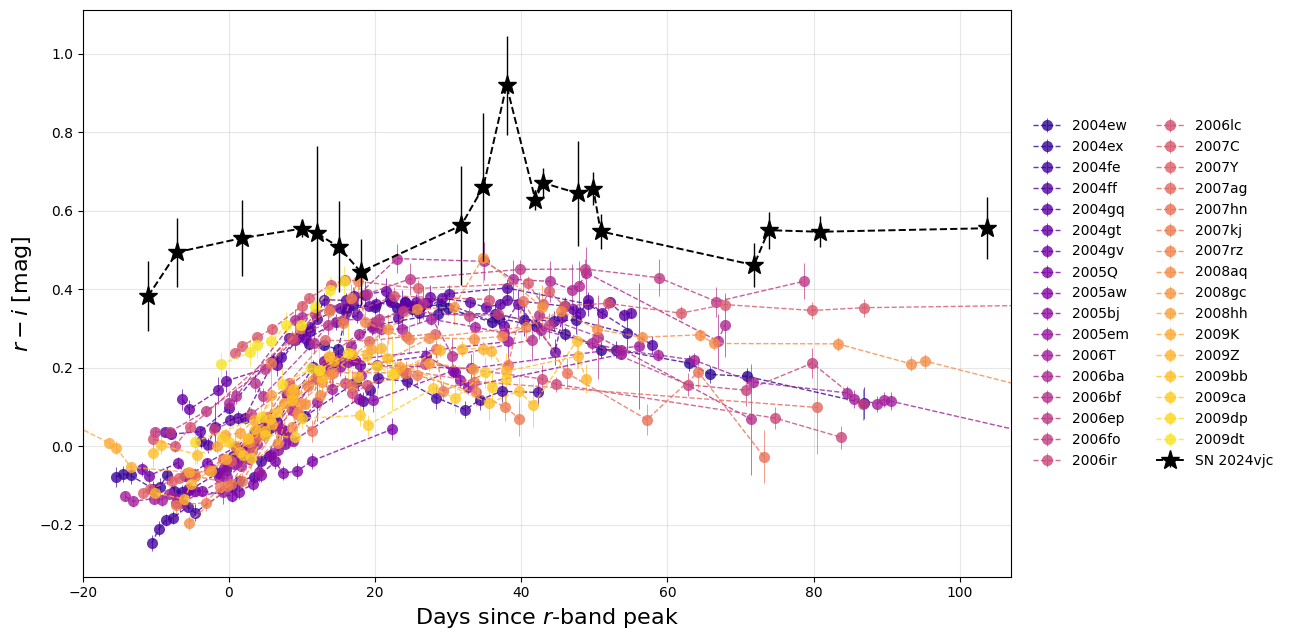}
\caption{Color-curve ($r - i$) evolution of SN 2024vjc (black stars) compared with sample from T18 (colored dots) as a function of days since $r$-band peak, only the MW extinction is corrected for. A host-extinction-corrected comparison is presented in Figure~\ref{fig:T18_color_curve_comparison}.}
\label{fig:color_curve}
\end{figure}

Figure~\ref{fig:color_curve} compares the $r-i$ color evolution of SN~2024vjc with the T18 sample. As with the light-curve comparison, we use T18 color curves that are corrected only for Galactic extinction. SN~2024vjc is systematically redder than the comparison objects at all epochs. Around maximum light, the color reaches $r-i \approx 0.4$ mag, already exceeding the bulk of the sample, and it evolves to values $\gtrsim 0.6$ mag at later phases. The light curve and color evolution therefore point toward substantial host attenuation; assuming 
$E(r-i)$ $\sim$ $0.4$ mag, this corresponds to $E(B-V)_{host}$ $\sim$ $0.6$ mag. This is consistent with the host attenuation estimates derived from independent diagnostics presented in Section~\ref{sec:attenuation}.


We now return to the small excess noted above in the $i$- and $z$-band light curves around MJD $\sim$60\,625. We associate this feature with the emergence of nebular emission features at this epoch, specifically the Ca~\textsc{ii} NIR triplet ($\lambda\lambda 8498,8542,8662$; falling within $z$) and the [Ca~\textsc{ii}] $\lambda\lambda 7291,7324$ doublet (falling within $i$), which enter the broadband filters as SN~2024vjc transitions to the nebular phase. A qualitatively similar redward excursion in $r-i$ near this phase is also seen in a preliminary comparison with SN~2007hn, SN~2006lc, and SN~2006fo (Type~Ib; T18; \citealt{2014ApJS..213...19B}), SN~2006lv (Type~Ib; \citealt{2014ApJS..213...19B}), SN~2020akf (Type~Ic; \citealt{2026arXiv260712525K}), SN~2019yvr (Type~Ib; \citealt{2024MNRAS.529L..33F}), and SN~2020bvc (Type~Ic-BL; \citealt{2021ApJ...908..232R}), suggesting this may be a generic feature of the nebular transition; however, $i$-band coverage in the literature is sparse for most SESNe and rarely well sampled specifically through this phase window, so we do not attempt a quantitative assessment of how common or statistically significant this feature is therefore lies beyond the scope of the present work and is left for future investigation.

\section{Spectra}\label{sec:Spectra}

\begin{figure}[ht!]
\centering
\includegraphics[width=0.8\linewidth]{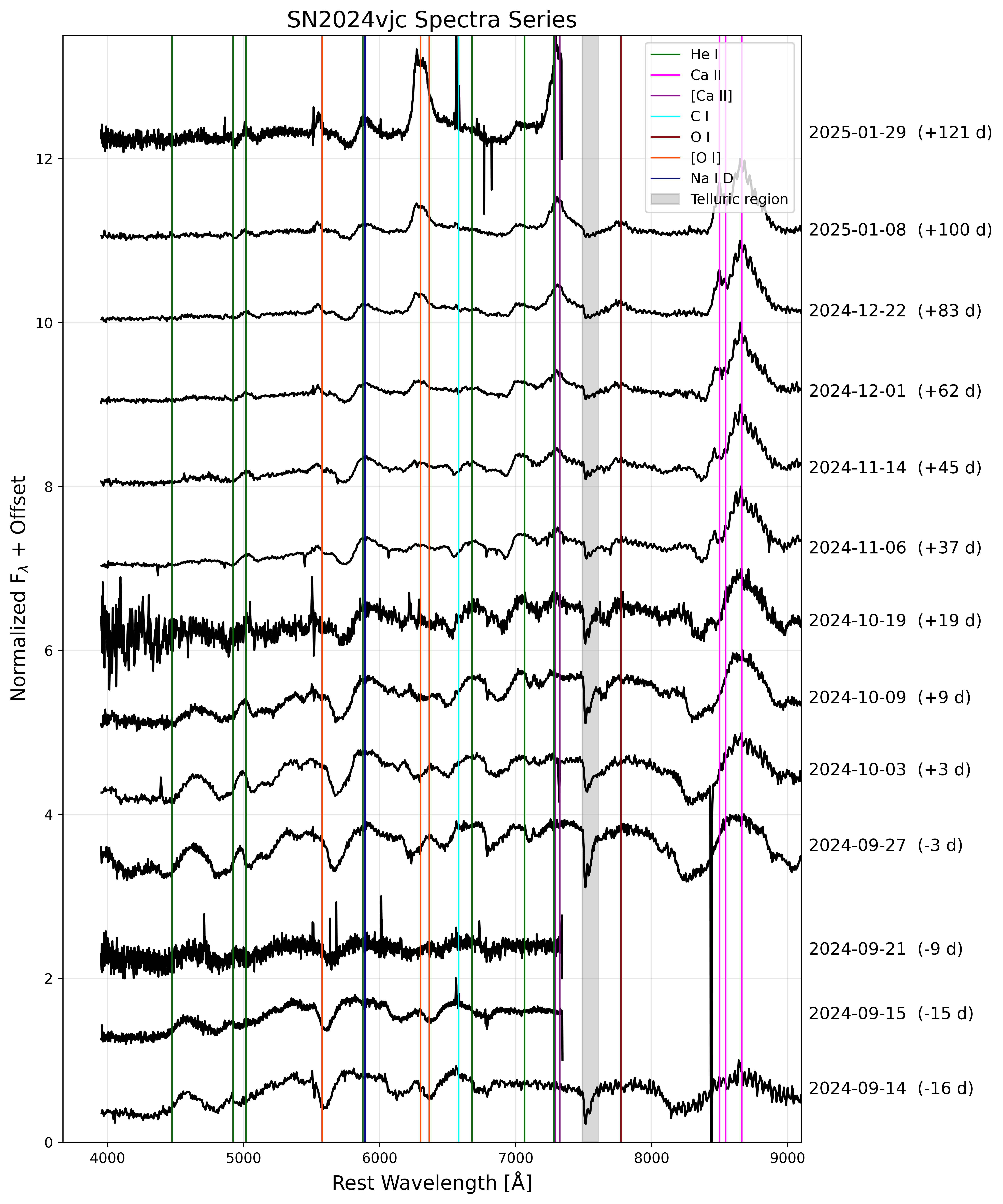}
\caption{Time series of the observed optical spectra of SN 2024vjc. The gray shaded region marks the telluric O$_2$ A-band at $\sim$7600~\AA. Vertical colored lines indicate the rest-frame wavelengths of key spectral features; absorption minima appear blueshifted by the photospheric velocity (see \ref{fig:He_velocity} for the He $\lambda$5876 velocity evolution).}
\label{fig:spectra_timeseries}
\end{figure}

We present the optical spectra of SN 2024vjc in Figure \ref{fig:spectra_timeseries}, covering 11 epochs from $-14$~d (2024-09-14) up to $+123$~d (2025-01-29) relative to the $g$-band maximum. To ensure consistency in the early-time continuum, the $-15$~d GMOS-S spectrum was scaled to match the slope of the $-14$~d EFOSC2 reference as it originally showed a slightly bluer continuum slope than the rest of the sequence, likely due to an issue in the flux calibration. The telluric region at $\sim 7600$~\AA, corresponding to the O$_{2}$ A-band, is indicated by a gray area. 

The presence of He~I and the absence of H~lines confirm the Type~Ib classification. As the SN evolves, the permitted Ca~II NIR triplet ($\lambda\lambda$8498, 8542, 8662) and O~I $\lambda$7774 features increase in strength. Furthermore, the spectra show the emergence and subsequent strengthening of forbidden emission lines as the ejecta transition toward the nebular phase, most notably the [O~I] $\lambda\lambda$6300, 6364 and [Ca~II] $\lambda\lambda$7291, 7323 doublets.

\begin{figure}[ht!]
\centering
\includegraphics[width=0.9\linewidth]{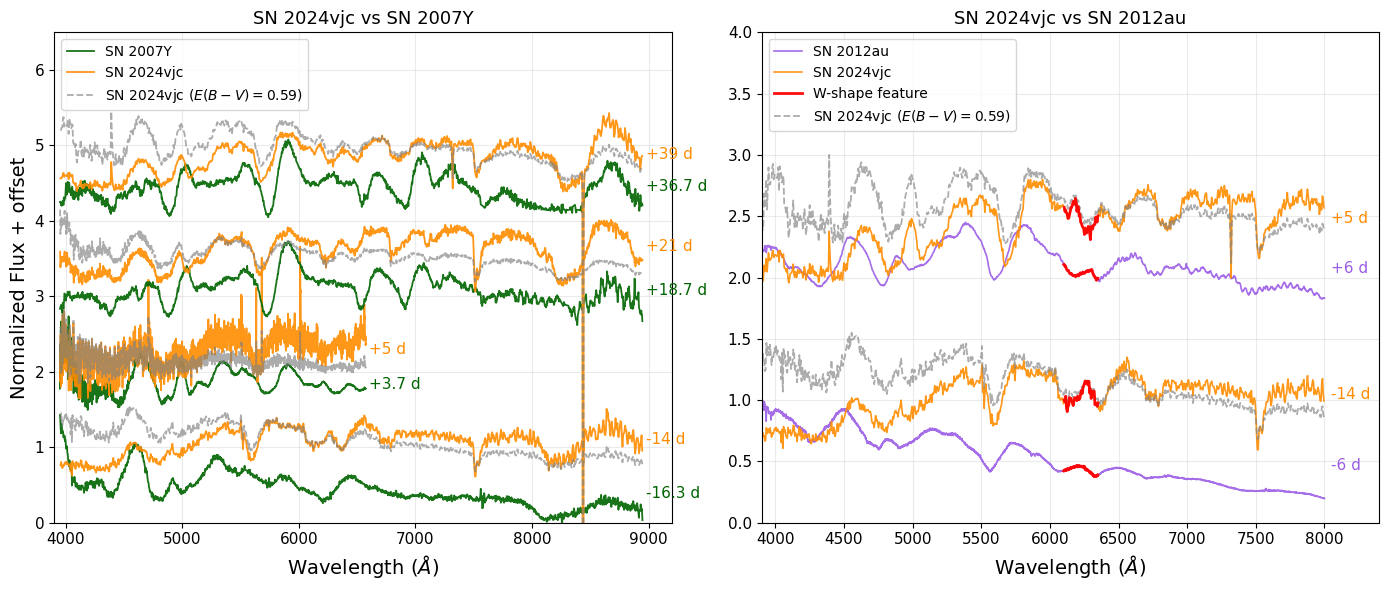}
\caption{Spectral comparison of SN~2024vjc against two reference SNe~Ib at similar phases. \textit{Left:} SN~2024vjc (orange) compared with the intrinsically faint SN~2007Y (green) at four matched epochs ($-14$, $+5$, $+21$, and $+39$~d). \textit{Right:} SN~2024vjc compared with SN~2012au (purple) at two matched early epochs ($-14$ and $+5$~d), with the W-shaped feature around $\sim 6300$~\AA\ shared by both objects highlighted in red. In both panels, gray dashed lines show SN~2024vjc corrected for host-galaxy extinction assuming $E(B-V){_{\rm host}}$ = 0.59 and a CCM89 extinction law with R$_{V}$ = 3.1. Spectra are normalized and vertically offset for clarity. Phases are quoted relative to $g$-band maximum.}
\label{fig:sn2007Y_SN2012au_spectra}
\end{figure}

To place SN~2024vjc in the context of the SESN diversity, we first compare it with SN~2007Y \citep{2009ApJ...696..713S}, a well-studied intrinsically faint SESN from the T18 sample. We note that the classification of SN~2007Y is somewhat ambiguous: while originally classified as Type~Ib by \citet{2009ApJ...696..713S}, the persistence of weak hydrogen features in its spectra has led some subsequent works to favor a Type~IIb classification \citep[see][]{2010MNRAS.409.1441M}. Regardless of the exact subtype, SN~2007Y is representative of the faint end of the SESN luminosity distribution, which is the relevant comparison for the question we address here. We further note that SN~2007Y shows some signs of interaction at late phases, making it somewhat peculiar; however, at the early epochs considered here these signatures had not yet developed, and the comparison remains valid.
Figure~\ref{fig:sn2007Y_SN2012au_spectra} shows the spectral evolution of both objects at four matched epochs ($-14$, $+5$, $+21$, and $+39$~d). The two SNe share broadly consistent spectral morphology. Some divergence in the relative line strengths emerges at the latest epoch ($+39$~d), which may reflect the onset of the interaction signatures known to develop in SN~2007Y at later phases.

A major difference between SN~2024vjc and other SNe~Ib is seen in the continuum slope and color, as was already indicated in the color curve analysis. The fact that SN~2024vjc appears redder than even an intrinsically faint object indicates that its red color cannot be explained by a low photospheric temperature alone. An additional reddening contribution, naturally provided by host-galaxy attenuation is required. To strengthen this argument, we compare in Figure~\ref{fig:sn2007Y_SN2012au_spectra} (right panel) a spectral sequence of SN~2024vjc with SN~Ib~2012au. This object was selected because both share the `W'-shaped feature around 6300~\AA, allowing a direct comparison of the spectral morphology while isolating the effect of the continuum slope. We note that the larger blueshifts and line widths seen in SN~2012au can be attributed to its more energetic nature \citep{2013ApJ...770L..38M} related to a central engine powering. SN~2024vjc is also redder than SN~2012au in the earliest spectra. The fact that SN~2024vjc is redder than both an intrinsically faint and an intrinsically luminous SN~Ib strongly supports the presence of substantial host-galaxy attenuation rather than an intrinsic photospheric origin.

The `W'-shaped feature is sometimes seen in SNe Ib and it may arise from high velocity H$\alpha$ or Si II $\lambda$6355 \citep{2015ApJ...815..120M}. A morphologically similar, though not necessarily physically connected, feature has been widely discussed in hydrogen-rich SNe (SNe II), including early studies of objects such as SN 1999em and SN 2005cs (e.g., \citealt{2000ApJ...545..444B, 2006MNRAS.370.1752P}). \citet{2017ApJ...850...89G} systematically investigated this feature, naming it ``Cachito'' and showing that it is present in $\sim$60$\%$ of their SN II sample. For SN~2024vjc, our \textsc{SYN++} modelling (Section~\ref{sec:SYN++_lower_temp}) provides a significantly better match to this feature with Si~{\sc ii} $\lambda$6355, suggesting that while the `W'-shaped feature may look morphologically similar across different objects, its physical origin can differ from case to case.

\begin{figure}[ht!]
\centering
\includegraphics[width=0.6\linewidth]{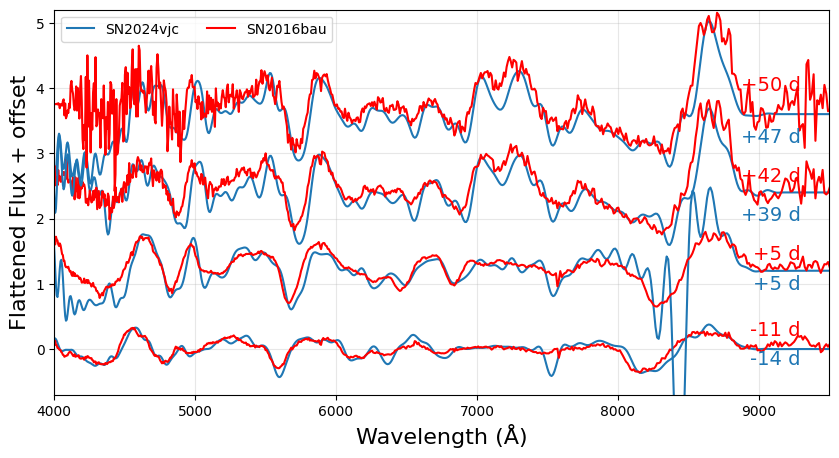}
\caption{The flattened spectra of SN 2024vjc (blue) at four epochs, compared with those of SN 2016bau (red) at similar phases.}
\label{fig:sn16bau_SNID_spectra}
\end{figure}

To further characterize the intrinsic spectral properties of SN~2024vjc independently of the continuum slope, and to compare against a more representative sample of normal Type~Ib SNe than the two cases discussed above, we compare flattened spectra following the \textsc{SNID} routines \citep{2007ApJ...666.1024B,2011ascl.soft07001B}, enabling a direct comparison with young Type~Ib SNe from \citet{2025A&A...693A.307Y}. Among the objects in that sample, SN~2016bau shows the closest spectral match to SN~2024vjc at multiple epochs. Figure~\ref{fig:sn16bau_SNID_spectra} shows the spectral evolution of both objects at similar epochs; the spectra are remarkably similar at all phases, with the only notable difference being the presence of a `W'-shaped feature around 6300~\AA\ in SN~2024vjc (see above).

In summary, the spectral morphology and evolution of SN~2024vjc is consistent with standard Type~Ib SNe, as demonstrated by the flattened-spectrum comparison with SN~2016bau, while its continuum slope is systematically redder than both intrinsically faint (SN~2007Y) and energetic (SN~2012au) reference objects. The spectral properties therefore point to the same question raised by the photospheric analysis $-$ whether the red continuum arises from an intrinsically low photospheric temperature or from substantial host-galaxy attenuation.

\subsection{He~I \texorpdfstring{$\lambda$5876}{λ5876} velocity evolution}

\begin{figure}[ht!]
\centering
\includegraphics[width=0.48\linewidth]{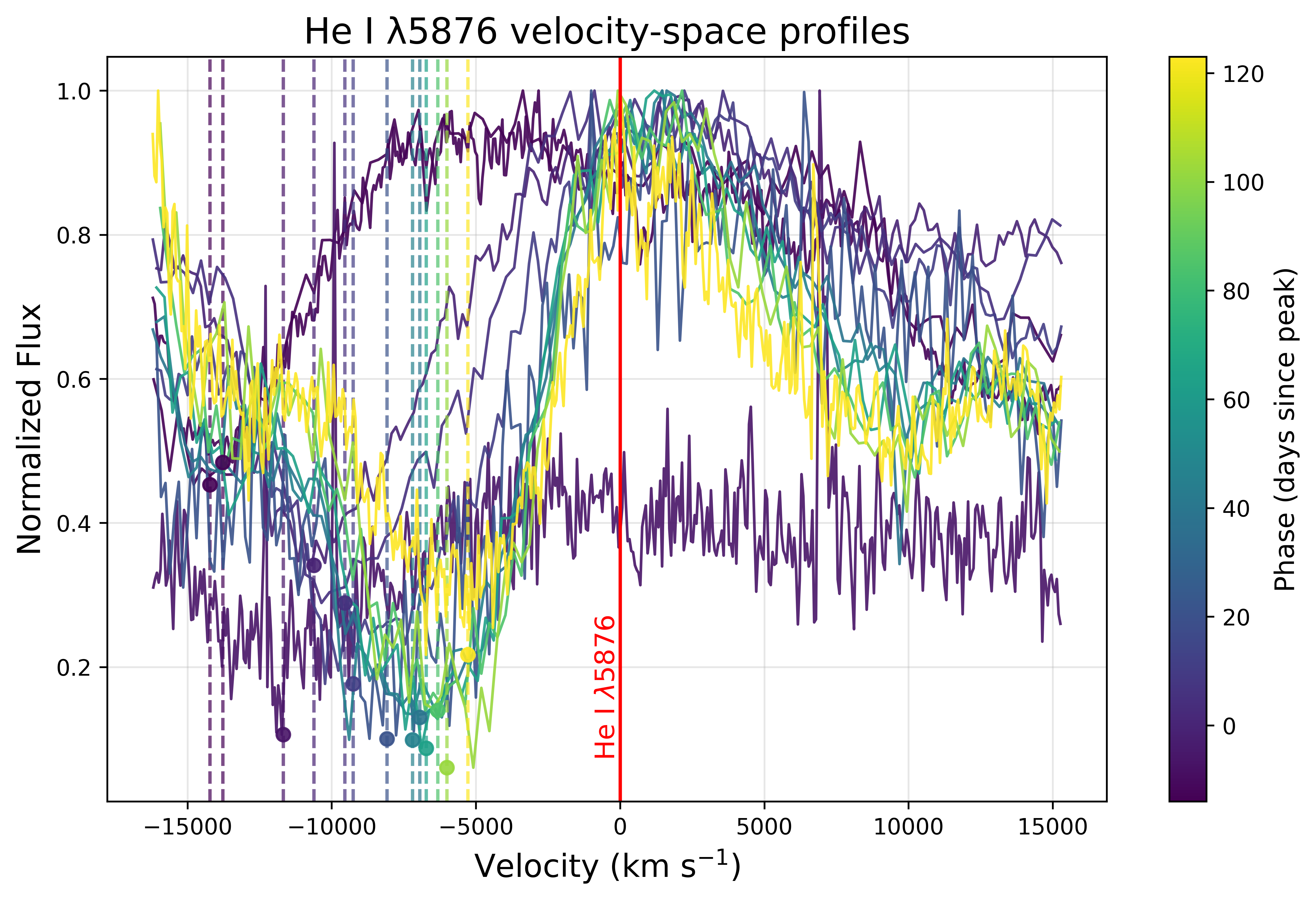}
\includegraphics[width=0.46\linewidth]{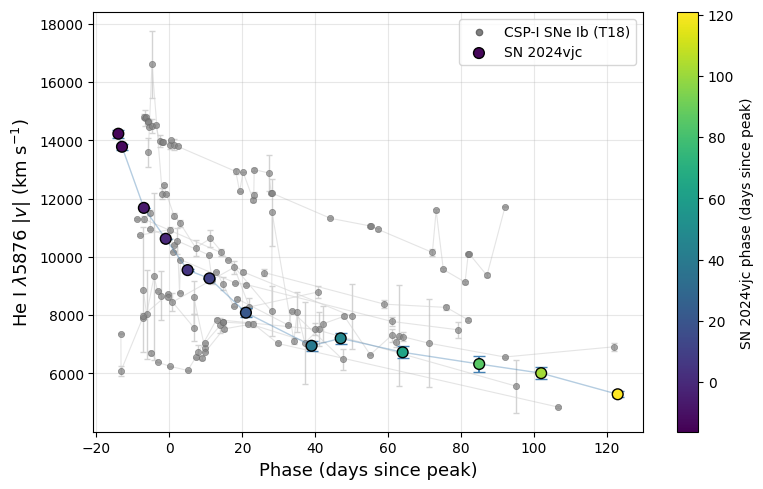}
\caption{\textit{Left:} Velocity space profiles of the He I $\lambda$5876 line (the red vertical line indicates the rest wavelength). Spectra are color coded by phase relative to peak brightness; the colored-dotted lines mark the absorption minima determined from the polynomial fits. \textit{Right:} Temporal evolution of the He I $\lambda$5876 absorption velocity. Gray points show the CSP-I Type~Ib sample from T18, while filled circles show SN~2024vjc color-coded by phase.}
\label{fig:He_velocity}
\end{figure}

Figure \ref{fig:He_velocity} shows the evolution of the He I $\lambda$5876 absorption velocity in SN 2024vjc alongside the CSP-I Type~Ib sample from T18.
The velocity decreases from an initial $\sim$ 14\,000 km s$^{-1}$ to $\sim$ 6000 km s$^{-1}$ over the observed period (-16 to 121 days). This places SN 2024vjc at the high velocity end of the He I $\lambda$5876 distribution, while still within the range observed for normal Type Ib SNe, consistent with the broad diversity in He I velocities reported in previous studies \citep{2019MNRAS.485.1559P, 2023A&A...675A..83H}

\subsection{\textsc{SYN++} and lower temperatures limits} \label{sec:SYN++_lower_temp}

To identify the ions contributing to the spectral features of SN 2024vjc, we use the parametric spectral synthesis code \textsc{SYN++} \citep{2011PASP..123..237T}. The model assumes a blackbody continuum with superimposed, Doppler-broadened line profiles from user specified ions. The spectra of SN 2024vjc are significantly redder than those of typical Type Ib SNe, which may be due to dust attenuation or an intrinsically low photospheric temperature. To minimize the impact of this effect on ion identification, we initially generate a flattened synthetic spectrum using \textsc{SYN++} and compare it with the flattened earliest observed spectrum of SN 2024vjc. This approach allows us to focus on relative line features rather than the overall spectral slope.

\begin{figure}[ht!]
\centering
\includegraphics[width=0.55\linewidth]{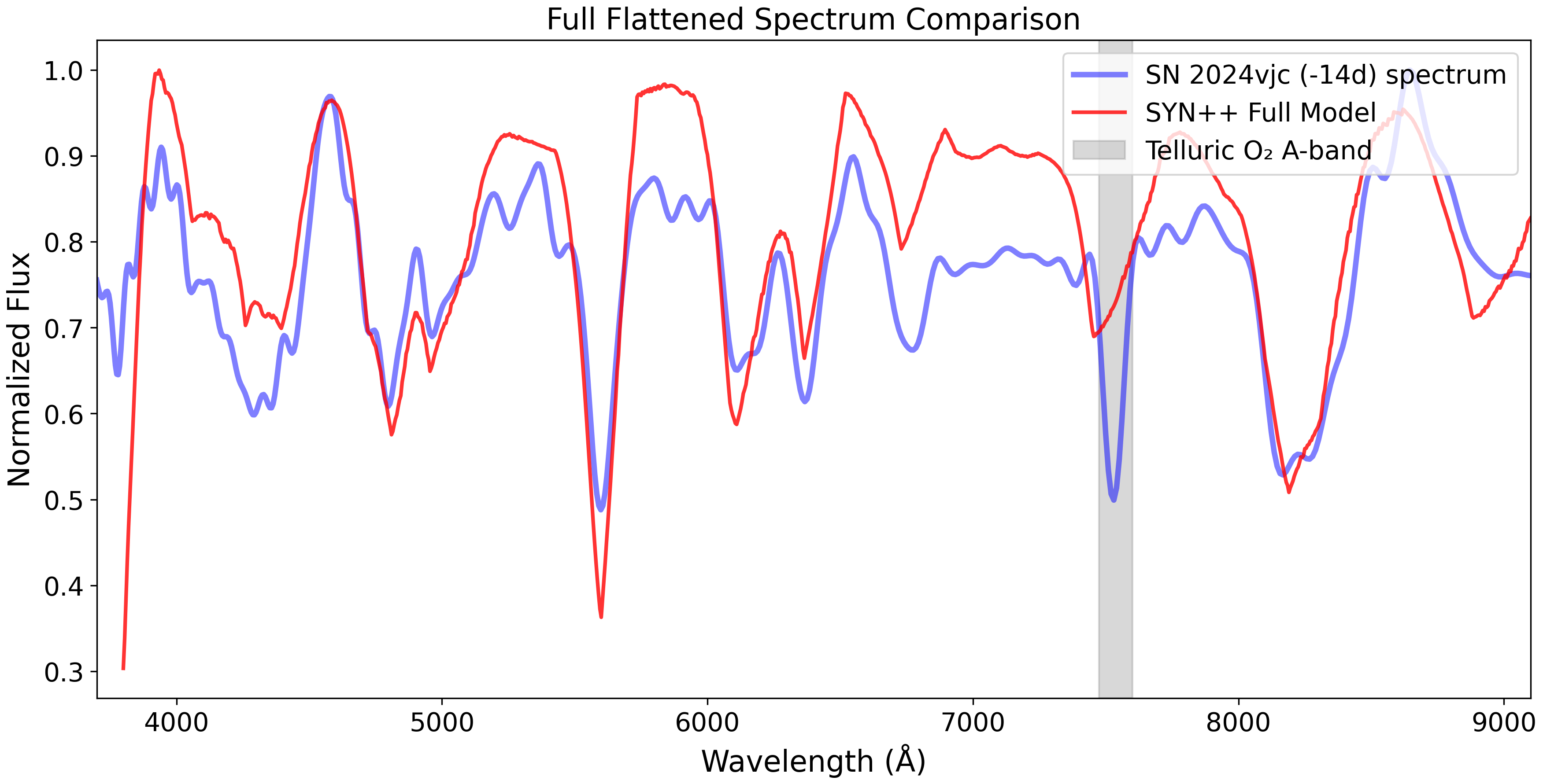}
\includegraphics[width=0.55\linewidth]{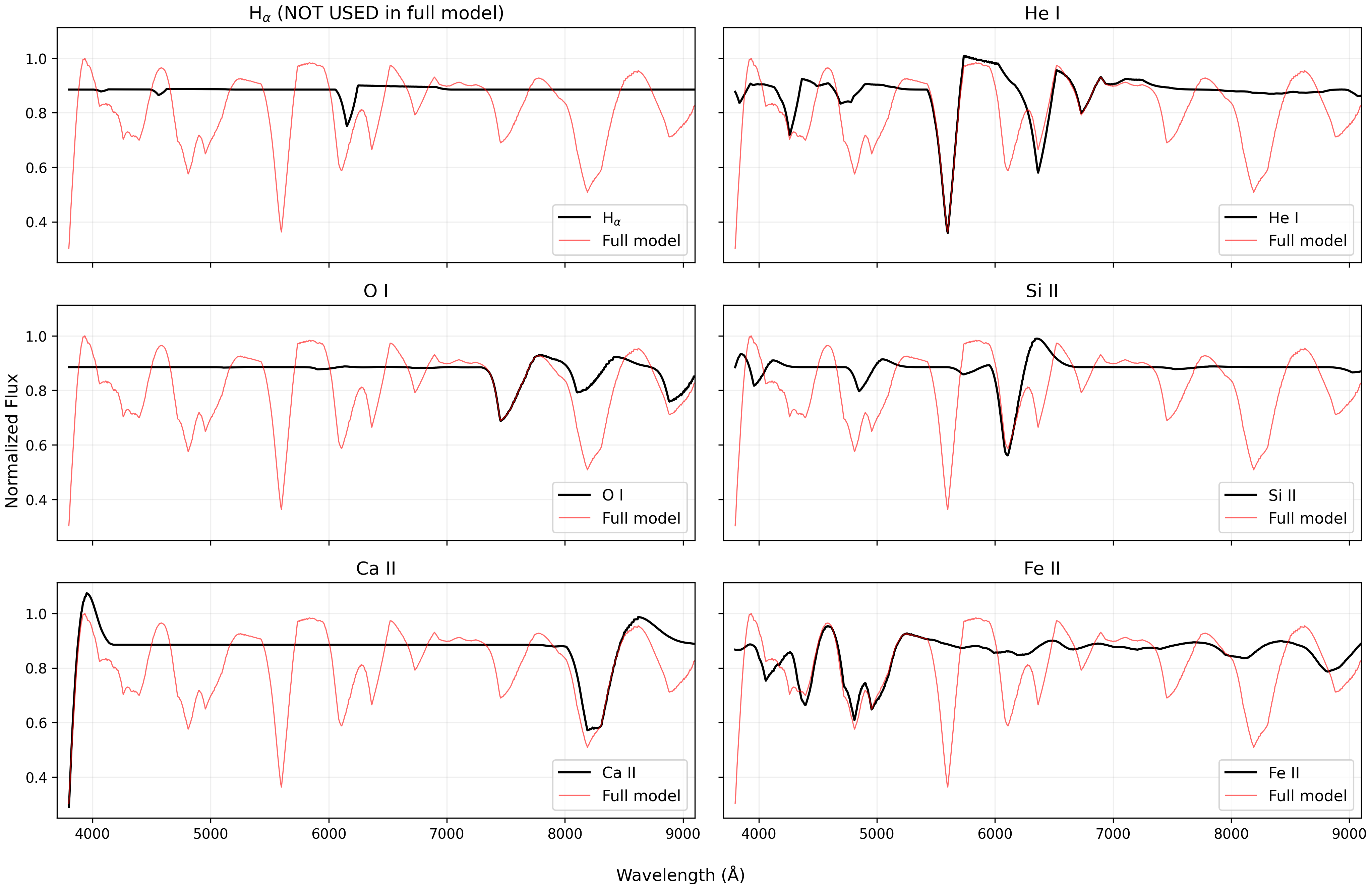}

\caption{Top: \textsc{SYN++} synthetic spectrum (red) compared with the flattened earliest spectrum for SN 2024vjc (blue). Ions used in this simulation are: He I, O I, Si II, Ca II, and Fe II.
Bottom: Contribution of each ion (black) to the \textsc{SYN++} final spectrum (red).}
\label{fig:SYN++_results}
\end{figure}

We adopt a \textsc{SYN++} configuration guided by the modeling presented by \citet{2013ApJ...770L..38M} for SN~2012au, given the similarity between the early-time spectra of the two objects. Specifically, we use the same set of ions included in their model as a starting point, while the photospheric velocity, temperature, and individual ion parameters (optical depth, excitation temperature, and velocity range) are independently adjusted to best match the observed spectrum of SN~2024vjc. The ions included in the final model are He~{\sc i}, O~{\sc i}, Si~{\sc ii}, Ca~{\sc ii}, and Fe~{\sc ii}.We also include high-velocity H$\alpha$ to investigate the origin of the 6300~\AA\ absorption feature, which in some stripped-envelope SNe has been attributed to trace amounts of hydrogen. However, in the case of SN 2024vjc, we find that Si~{\sc ii} $\lambda$6355 provides a significantly better match, and H$\alpha$ is therefore not included in the final model.  After optimizing the \textsc{SYN++} parameters, Figure~\ref{fig:SYN++_results} presents the final \textsc{SYN++} fit to the observed spectrum, along with contributions from each ion included in the model. Although H$\alpha$ was not used in the simulation, we show its theoretical profile for comparison with the Si II features.

Once the ions responsible for the observed spectral features are identified using the flattened spectra, we compare the non-flattened \textsc{SYN++} synthetic spectra with the observed SN 2024vjc spectrum at $-14$ d for different values of the photospheric temperature. 
Here, no  host-galaxy  attenuation is assumed - therefore, the temperature here provides the lower limit on the blackbody temperature associated with the SN. 

\begin{figure}[ht!]
\centering
\includegraphics[width=0.60\linewidth]{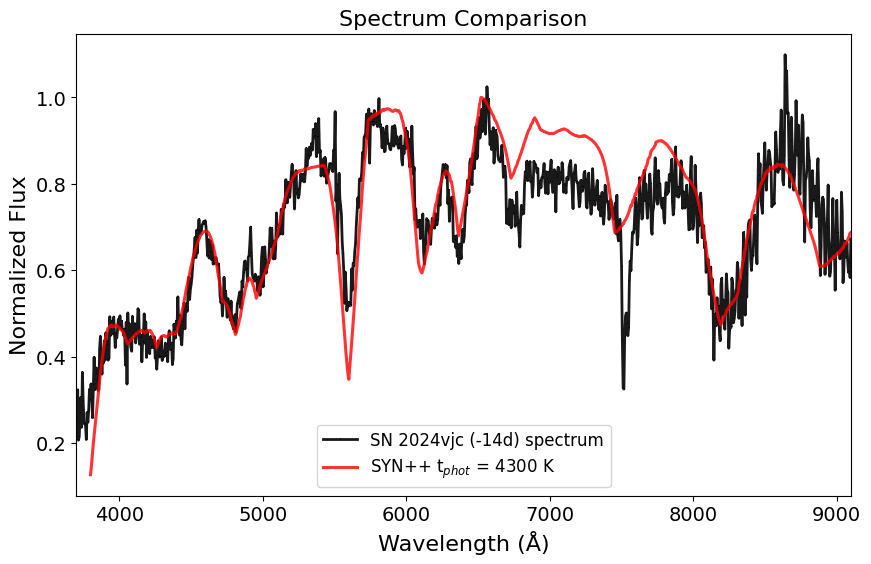}

\caption{Comparison of the observed -14 day spectrum of SN 2024vjc (black) with \textsc{SYN++} synthetic spectra for photospheric temperatures of 4300 K (red). No host extinction is assumed in this figure.}
\label{fig:SYN++_temperature_limits}
\end{figure}

The synthetic spectrum at $T_{\mathrm{phot}} = 4300$\,K provides a good match to the observed continuum and spectral features of SN~2024vjc at $-14$\,d, when no host-galaxy attenuation is assumed (Figure~\ref{fig:SYN++_temperature_limits}). However, this temperature is significantly lower than the blackbody temperatures typically observed in SNe~Ib; the CSP-I Type~Ib sample from T18 shows blackbody temperatures of $\gtrsim 6000$\,K even for their coolest objects (see Section \ref{sec:pseudo-bolometric} and Figure~\ref{fig:bol_lc}). We note that in this regime (pre-maximum light) $T_{\mathrm{phot}} \approx T_{\mathrm{BB}}$, given that the ejecta are optically thick and the SED is well-approximated by a Planck function at these early epochs.
Consequently, achieving a physically realistic photospheric temperature requires the inclusion of an additional reddening component. We note that the 4300\,K value derived under the no-attenuation assumption represents a lower limit on the intrinsic photospheric temperature of SN~2024vjc. 

There is another argument against the intrinsically low temperature, based on the SYN++ fit that constrains the excitation temperature ($T_{\mathrm{exc}}$) to reproduce the relative line strength ratios. We find $T_{\mathrm{exc}} \sim 10\,000$ K, much higher than the blackbody temperature ($\lesssim$5000 K) that is required to roughly explain the continuum color without the host extinction. In particular, the Ca~{\sc ii} NIR triplet ($\lambda$$\lambda$8498, 8542, 8662) weakens as $T_{\mathrm{exc}}$ drops, becoming inconsistent with the observed spectrum at those wavelengths. This discrepancy implies that the intrinsic photosphere is hotter than suggested by the spectral slope, supporting dust attenuation as the primary cause of the red continuum.

\section{Attenuation toward SN 2024\MakeLowercase{vjc}} \label{sec:attenuation}

As demonstrated in the analyses in the previous sections, SN 2024vjc seems to behave as a normal SN Ib except for its exceptionally high reddening deduced from the LC and spectroscopic evolution. In this section we will further investigate this issue, by comparing the attenuation properties inferred from the SN observables with that obtained through the widely-used SN-independent method, i.e., the Na I D absorption depth. 


\subsection{Attenuation estimated from the SN observed properties}


To quantify the attenuation, we first assume that the intrinsic evolution of SN~2024vjc follows the standard template behaviors established for the SESNe population. Under this framework, any observed color excess is attributed to extrinsic reddening. 

\begin{figure}[ht!]
\centering
\includegraphics[width=0.7\linewidth]{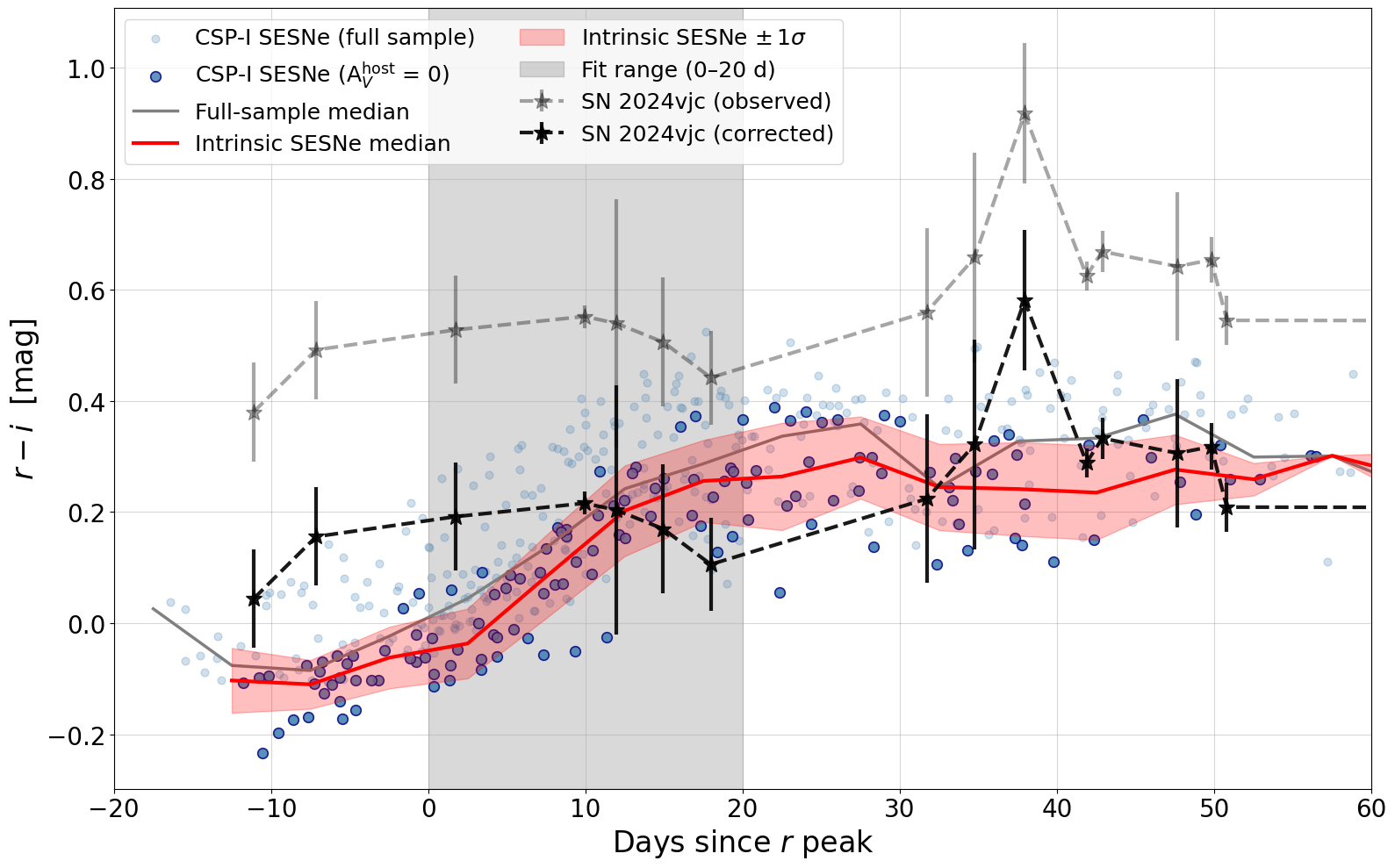}

\caption{$r-i$ color evolution of SN 2024vjc compared to the CSP-I SESNe sample of T18. Light blue circles show the full CSP-I sample, and dark blue circles highlight the subset of CSP-I SESNe with negligible host-galaxy attenuation ($A_V^{\mathrm{host}} = 0$). The gray solid line shows the median color curve of the full sample, while the red solid line and shaded band show the median and $1\sigma$ envelope of the intrinsic (no host attenuation) sub-sample, which we adopt as the reference for the color-excess measurement. SN 2024vjc's observed colors (gray stars) and the host-corrected colors (black stars) are obtained after applying $E(r-i) = 0.336$\,mag derived from the intrinsic SESNe reference curve. The gray shaded vertical band indicates the 0--20 d post-$r$-peak phase range used for the fit.}
\label{fig:T18_color_curve_comparison}
\end{figure}

Figure~\ref{fig:T18_color_curve_comparison} shows the observed and corrected $r-i$ color evolution of SN~2024vjc compared to the CSP-I SESNe sample and the corresponding reference curves. First, we use the intrinsic color-curve templates presented by \citet{2018A&A...609A.135S}. By comparing our observed photometry in the phase range from 0 to +20 days relative to peak with these templates, we estimate the host-galaxy color excess $E(X-Y)$$_{host}$, for filter combinations shared between this work and \citet{2018A&A...609A.135S}. Applying this method, we obtain $E(r-i)$$_{host}$ = 0.391 $\pm$ 0.294 mag, indicating that SN 2024vjc suffers from significant host-galaxy attenuation. To convert this value to the standard $E(B-V)$, we use the $A_{b}$/$E(B-V)$ coefficients from Table 6 from \citet{2011ApJ...737..103S} for the $i$ and $r$ bands assuming $R_V$ = 3.1 for the  host-galaxy . This yields  $E(B-V)_{host}$ = 0.667 $\pm$ 0.501 mag. We note that the large uncertainty on this estimate ($\pm 0.501$\,mag) makes it weakly constraining on its own. It is consistent with substantial host attenuation but does not pin down a precise value, and we therefore treat the more precise color-evolution estimate derived below as the 
primary photometric reference. 

Another approach to estimate the host-galaxy attenuation is based on the color-curve method inspired by \citet{2011ApJ...741...97D}, which relies on the relatively uniform intrinsic color evolution of SESNe between 0 and +20~days after maximum light. The observed color evolution is compared with a reference color curve to infer the color excess. In this work, we use the $r-i$ color curves of SESNe from the CSP-I sample presented by T18 as the reference set (Figure~\ref{fig:T18_color_curve_comparison}).

We construct two reference curves. The first uses the full CSP-I SESNe sample. Because this sample includes objects affected by host-galaxy attenuation, the resulting median color evolution is expected to be systematically redder than the true intrinsic SESNe trend, and the inferred color excess therefore represents a lower limit. Applying this method to SN~2024vjc over the phase range $0$--$20$~days relative to $r$-band maximum yields $E(r-i)_{\mathrm{lower}} = 0.303 \pm 0.076$~mag, corresponding to $E(B-V)_{\mathrm{host}} > 0.52$~mag for $R_V = 3.1$. The second reference curve is built using only CSP-I SESNe with negligible host attenuation, which we select as those objects with $A_V^{\rm host} = 0$ (or consistent with zero within uncertainties) in the sample presented by T18. This subsample provides a cleaner estimate of the intrinsic SESNe color evolution. In this case we obtain $E(r-i) = 0.336 \pm 0.054$~mag, corresponding to $E(B-V)_{\mathrm{host}} = 0.57 \pm 0.09$~mag.

We note that between 0 and +20~days the color evolution of SN~2024vjc appears flatter than the average SESNe trend, which typically shows steady reddening during this phase. This may indicate that SN~2024vjc has intrinsic color properties that deviate from standard templates, and increases the uncertainty in the derived color excess. The high $E(B-V)$ nonetheless remains consistent with the attenuation estimates derived from other independent diagnostics.

In summary, the comparison with the intrinsic color templates of \citet{2018A&A...609A.135S} gives $E(B-V)$ = 0.667 $\pm$ 0.501 mag, while the color curve fitting method based on \citet{2011ApJ...741...97D} yields $E(B-V)$ = 0.57 $\pm$ 0.09 mag. The photometric methods thus consistently point to substantial attenuation toward SN 2024vjc. Alternatively, we may use the full spectral window to constrain the attenuation properties as a continuous function of wavelength, using a high signal-to-noise spectrum. In other words,  we may constrain $E(B-V)$ and $R_{V}$ simultaneously if we adopt the CCM89 attenuation curve. We note that spectral dereddening has been successfully employed in other contexts; for example, \citet{2000ApJ...537..861S} derived the attenuation toward the Crab Pulsar by fitting a power-law to its dereddened spectrum. For SNe, the intrinsic spectral shape must instead be inferred from a comparison object with known attenuation properties.
 

We de-redden the earliest spectrum of SN 2024vjc using a CCM89 attenuation curve, exploring a grid of attenuation parameters. Prior to minimization of the sum of squared residuals (SSR) (see below), both spectra are independently normalized by their peak flux. As a consequence of this normalization, the absolute flux offset between the two spectra - which would be required to independently constrain $A_{V}$ - is removed. The method therefore effectively has one free parameter, $E(B-V)=A_{V}/R_{V}$, rather than two independent parameters. The grid search over $A_{V}^{*}$ and $R_{V}^{*}$ (where the asterisks denote that these are apparent, normalization-dependent quantities rather than the true physical $A_{V}$ and $R_{V}$) is nonetheless performed, as it allows us to map the full degeneracy ridge and derive the uncertainty on $E(B-V)$ from the spread of values within the $\Delta$SSR$\leq$1 contour. We use a grid of  $A_{V}^{*}$ going from 0.5 to 5 in steps of 0.1, and $R_{V}^{*}$ going from 2 to 6 in steps of 0.1. We quantify the agreement between the de-reddened SN 2024vjc spectrum and the comparison SN~2012au spectrum using the SSR,

\begin{equation}
\mathrm{SSR} = \sum_{i} \left[ F_{\mathrm{24vjc,\,dered}}(i) - F_{\mathrm{SN Ib}}(i) \right]^2
\end{equation}
where $F_{\mathrm{24vjc,dered}}(i)$ is the peak-normalized, de-reddened flux of SN 2024vjc at the $i$-th wavelength bin and $F_{\mathrm{SN,Ib}}(i)$ is the peak-normalized flux of the comparison Type~Ib SN at the same wavelength. Lower values of SSR indicate better agreement between the spectral shapes

\begin{figure}[ht!]
\centering
\includegraphics[width=0.9\linewidth]{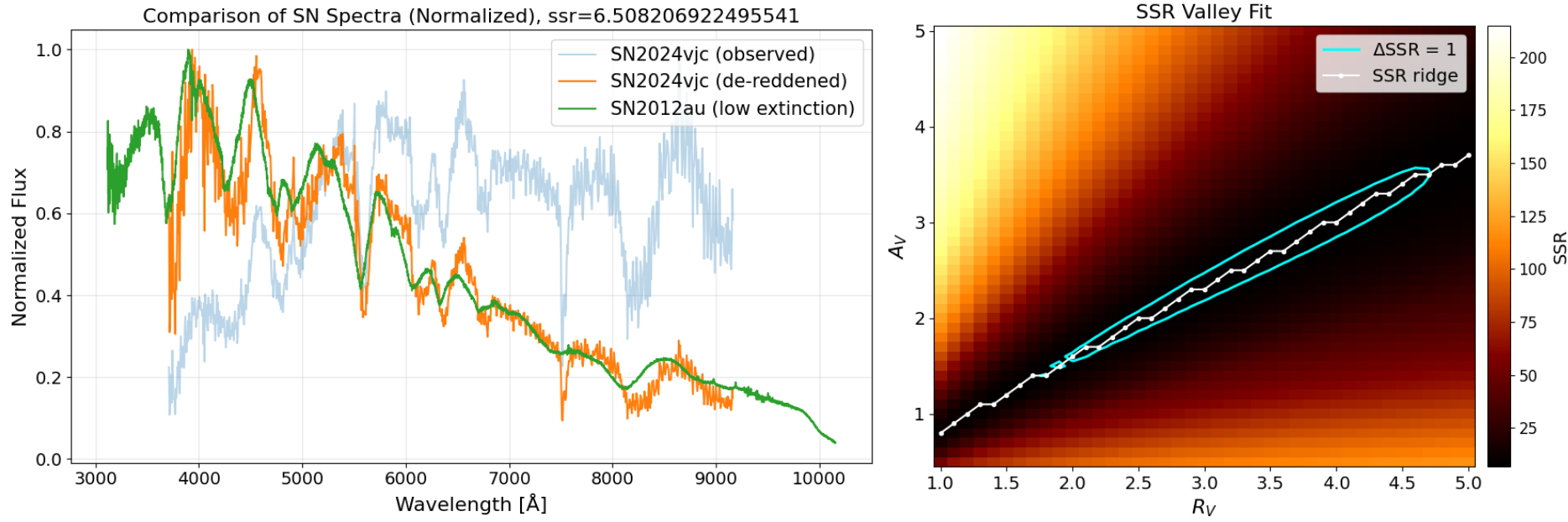}
\caption{\textit{Left panel:} Comparison of the earliest available spectra of SN 2024vjc and SN 2012au. The observed spectrum of SN 2024vjc is shown in semi-transparent blue, while the best fitting de-reddened spectrum is shown in orange. The earliest spectrum of SN 2012au, which suffers from minimal attenuation, is shown in green for comparison. \textit{Right panel:} Two-dimensional map of the sum of squared residuals (SSR) computed over a grid of $A_{V}$ and $R_{V}$ values. Darker regions indicate lower SSR values and therefore better spectral matches. The white curve traces the minimum-SSR ridge. The cyan contour encloses the region satisfying $\Delta$SSR$\leq$1.}
\label{fig:unreddened_spectrum}
\end{figure}

We compare SN 2024vjc with SN 2012au, which is previously shown to exhibit a similar spectral morphology at early epochs and is known to suffer little to no host-galaxy attenuation. The best fitting solution ($A_{V}^{*} = 2.6$ and $R_{V}^{*} = 3.4$) is shown in the left panel of Figure \ref{fig:unreddened_spectrum}, together with the original reddened spectrum for comparison. To estimate the color excess, we locate the minimum of the sum of SSR in the $A_{V}^{*}- R_{V}^{*}$ plane, which reveals the intrinsic degeneracy between these parameters. We define the statistically acceptable region as that satisfying $\Delta$SSR $<$ 1 relative to the global minimum (cyan contour in the right panel of Figure \ref{fig:unreddened_spectrum}). The color excess is then computed as $E(B-V)$~=~ $A_{V}^{*}/R_{V}^{*}$ along this region, with its uncertainty derived from the spread of values within the $\Delta$SSR $<$ 1 contour. This yields $E(B-V)$ = 0.767$_{-0.032}^{+0.033}$ mag. We note that this uncertainty is statistical only and does not capture the systematic from the choice of comparison object, which is difficult to quantify.

\subsection{Na I D absorption lines} \label{sec:Na_Id_attenuation}

With the attenuation properties derived on the basis of the SN observables, we now turn to the SN-independent method. The most straightforward and widely used method to estimate the host-galaxy color excess, $E(B–V)$, is to measure the EW of the Na I D absorption lines and apply the empirical relation of \citet{2012MNRAS.426.1465P}. However, in almost all of the spectra of SN 2024vjc we do not detect any Na I D absorption features. This non detection does not necessarily imply the absence of Na I D, as the spectral resolution and signal-to-noise ratio may be insufficient to resolve weak and narrow features. We therefore focus on estimating upper limits on the narrow Na I D EW that would be detectable in our data.

We follow the formalism presented by \citet{2001PASP..113..920L}. The $3\sigma$ equivalent width upper limit is calculated as $W_{\lambda}(3\sigma) = 3 \Delta\lambda \Delta I$, where $\Delta I$ is the $1\sigma$ root mean square fluctuation of the flux around a normalized continuum level. For the resolution element $\Delta\lambda$, we adopt the actual instrumental resolution (FWHM$_{\rm inst}$). These values, which range from $\sim 6.3$~\AA\ to $\sim 34.5$~\AA\ depending on the instrument and slit configuration, are detailed in Table \ref{tab:spectra}. For the continuum normalization, we applied a polynomial fit (ranging from 1st to 3rd degree depending on the local slope) to the regions flanking the expected Na~I~D position.

The upper limits for the EWs of Na~I~D absorption are listed in Table \ref{tab:ew_limits}. The upper limits generally increase toward later epochs as the signal-to-noise ratio decreases with the fading of the SN continuum, and do not reflect any physical evolution of the interstellar Na~{\sc i}~D absorption. 
\citet{2024A&A...687A.108G} showed that apparent variations in the EW of narrow ISM lines such as Na I D in low and mid-resolution SN spectra can arise from a combination of factors, including signal-to-noise, spectral resolution, slit configuration, and interference with the broad SN lines, rather than from intrinsic evolution of the absorbing material. In our case, the most plausible driver of the non-detections at later epochs is the declining S/N as the SN fades, since the Na I D feature is detected only in our highest-S/N spectrum (the $-$13 d GMOS-S spectrum). A genuinely time-varying Na\,\textsc{i}\,D absorption due to recombination of circumstellar Na\,\textsc{ii} to Na\,\textsc{i}, as observed in ILRTs by \citet{2023MNRAS.524.2978B}, would be expected to strengthen the line at later epochs; the opposite trend in our data, with the line detected only in the earliest high-S/N spectrum, may disfavor this scenario for SN 2024vjc. We therefore derive the most constraining 3$\sigma$ upper limit from the +21 d ALFOSC spectrum, yielding EW(${\rm Na~I~D}$) $<$ $0.77$~\AA. 

The empirical relation between Na~I~D EW and color excess $E(B-V)$ is most reliable in the regime below 1~\AA. 
We thus adopt EW(Na I D)$\lesssim$0.8~\AA\ as a representative 3$\sigma$ upper limit for non-detections. This value corresponds to a host-galaxy color excess of $E(B-V) \lesssim$ $0.20$ mag using \citet{2012MNRAS.426.1465P} relation.

\begin{deluxetable}{cc}
\tablecaption{Equivalent Width of Na I D Upper 3$\sigma$Limits for SN 2024vjc\label{tab:ew_limits}}
\tablehead{
\colhead{Phase (days)} & \colhead{EW Upper Limit (\AA)}}
\startdata
-14 & 1.81 \\
-7 & 7.94 \\
-1 & 1.88 \\
+5   & 2.36 \\
+11   & 1.68 \\
+21 & 0.77  \\
+39  & 2.55 \\
+47  & 2.55 \\
+63  & 3.31 \\
+85  & 2.85 \\
+102 & 3.28 \\
+123 & 1.5 \\
\enddata
\tablecomments{The relatively large value at $-7$\,d reflects a lower signal-to-noise ratio in that spectrum around 5890\,\AA\ rather than a true physical evolution.}
\end{deluxetable}

\begin{figure}[ht!]
\centering
\includegraphics[width=0.6\linewidth]{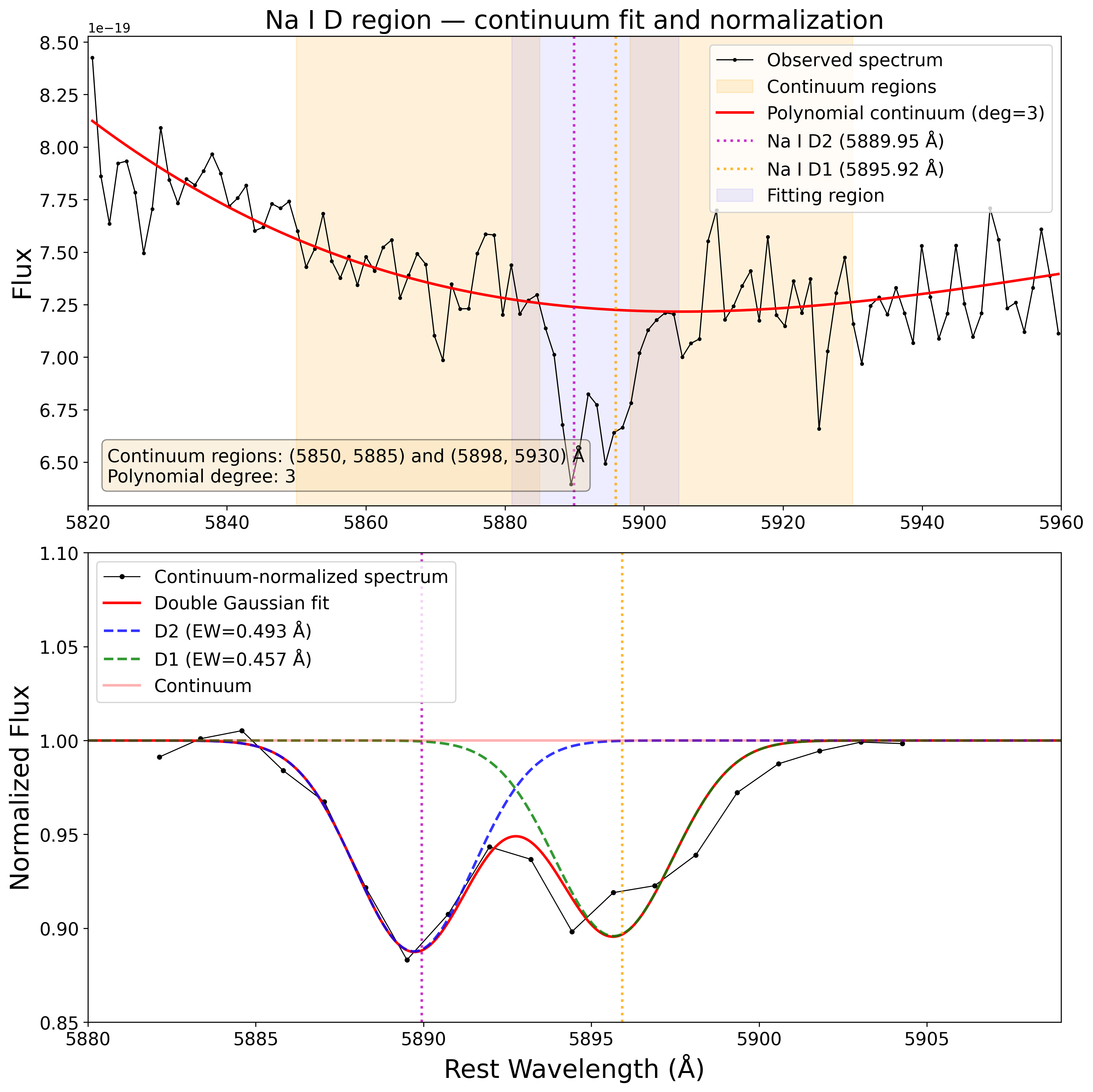}

\caption{\textit{Top:} Observed GMOS-S spectrum of SN~2024vjc around the Na~{\sc i}~D region ($5820$--$5960$~\AA). The red curve shows the degree-3 polynomial continuum fit, derived from the flanking regions $5850$--$5885$~\AA\ and $5898$--$5930$~\AA\ (shaded orange). The blue shaded region marks the fitting window used for the double-Gaussian model. Vertical dotted lines indicate the rest wavelengths of Na~{\sc i}~D2 ($5889.95$~\AA, magenta) and D1 ($5895.92$~\AA, orange).
\textit{Bottom:} Continuum-normalized spectrum (black) with the best-fit double-Gaussian model (red). The individual D2 and D1 components are shown as dashed blue and green curves, respectively. The fitted equivalent widths are EW(D2)~$= 0.493$~\AA\ and EW(D1)~$= 0.457$~\AA.}
\label{fig:EW_det}
\end{figure}

We marginally detect interstellar Na I D absorption in the GMOS-S spectrum obtained at $-$13 days relative to peak (Figure~\ref{fig:EW_det}). The upper panel of Figure~\ref{fig:EW_det} shows the observed spectrum over a wider wavelength range ($5820$--$5960$~\AA), together with the degree-3 polynomial continuum fit derived from the surrounding regions on either side of the doublet. We fit the feature with a double-Gaussian model, yielding equivalent widths of EW(D2) = 0.493 $\pm$ 0.062~\AA~and EW(D1) = 0.457 $\pm$ 0.059~\AA, for a total EW(D1+D2) = 0.949 $\pm$ 0.086 \AA.

The measured doublet ratio EW(D2)/EW(D1) $\approx$ 1 differs from the optically-thin value of 2 expected from the ratio of oscillator strengths \citep{2003ApJS..149..205M}. This could indicate saturation of the Na I D lines, or it could result from the narrow Na I D feature being superposed on the broad SN spectral structure, likely the He I $\lambda$5876 line an effect analogous to that shown by \citet{2024A&A...687A.108G} for narrow ISM lines superposed on broad SN features. In either case, the value reported here should be treated with caution. As this is nonetheless our only detection of the Na I D feature across the spectral series, we adopt it in the following discussion while bearing this caveat in mind.

Using the relations of \citet{2012MNRAS.426.1465P}, these values correspond to host-galaxy attenuation estimates of $E(B-V)$$_{\rm host}$ $\sim$ 0.14$^{+0.08}_{-0.05}$ mag from D2, 0.23$^{+0.16}_{-0.10}$ mag from D1, and 0.18$^{+0.06}_{-0.05}$ mag from the combined doublet, yielding consistent results irrespective of a line used in the analysis. Adopting the combined-line calibration, we infer $E(B-V)$$_{\rm host}$ $\sim$ 0.18 mag, which implies $A_{V}^{\rm host}$ $\sim$ 0.56 mag for $R_V$ = 3.1. 

We note that alternative empirical Na~{\sc i}~D EW to $E(B-V)$ relations have been proposed in the literature \citep[e.g.,][]{1990A&A...237...79B, 1994AJ....107.1022R, 1997A&A...318..269M, 2003fthp.conf..200T}. In particular, \citet{2003fthp.conf..200T} identified a second empirical branch among SNe~Ia in which more heavily reddened objects follow the relation $E(B-V) = -0.04 + 0.51 \times \mathrm{EW(Na~I~D)}$. Applying this relation to our detection, EW(D1+D2) = $0.949 \pm 0.086$\AA , yields $E(B-V)=0.444\pm0.044$ mag. Using instead the most constraining upper limit of EW(Na~I~D) $\lesssim0.77$~\AA\ gives $E(B-V)\lesssim 0.35$ mag. Both values are substantially higher than the \citet{2012MNRAS.426.1465P} estimate of $\sim$0.18 mag, and the detection-based value is in remarkable agreement with the Balmer decrement result ($E(B-V) = 0.449 \pm 0.148$ mag). This illustrates that the choice of empirical calibration can have a significant impact on the inferred attenuation, and further motivates the use of multiple independent diagnostics.

\begin{deluxetable}{lr}
\tablecaption{SN 2024vjc  host-galaxy  $E(B-V)$ inferred values \label{tab:E(B-V)_values}}
\tablehead{
\colhead{Method} & \colhead{ host-galaxy  $E(B-V)$ [mag]}}
\startdata
Na I D EW \citep{2012MNRAS.426.1465P} & 0.18$^{+0.06}_{-0.05}$ \\
Na I D EW \citep{2003fthp.conf..200T} & 0.444 $\pm$ 0.044 \\
Balmer decrement (H$\alpha$/H$\beta$) & $0.449 \pm 0.148$ \\
Color Template (+10 d)  & 0.667 $\pm$ 0.501 \\
Color Evolution (0-20 d) & 0.59 $\pm$ 0.09 \\
Spectrum de-reddening  & 0.767$_{-0.032}^{+0.033}$ 
\enddata
\end{deluxetable}

\subsection{Balmer Decrement}
\label{sec:balmer}

An additional, SN-independent estimate of the host-galaxy attenuation can be obtained from the Balmer decrement of the underlying \ion{H}{2} region detected in the $+123$\,d GMOS-S spectrum (Section~\ref{sec:Disc_HII}). Assuming Case~B recombination value of H$\alpha$/H$\beta$ = 2.86, we derive $E(B-V)_{\rm Balmer} = 0.449 \pm 0.148$\,mag, following Equation 3 from \citet{2013ApJ...763..145D} with $k(\mathrm{H}\alpha) = 3.326$ and $k(\mathrm{H}\beta) = 4.598$. This value is somewhat lower than the attenuation estimates derived from the SN light-curve color evolution (while they agree within 1$\sigma$) and spectral comparison methods, but significantly exceeds the Na\,\textsc{i}\,D-based estimate. 

All the attenuation estimates are summarized in Table~\ref{tab:E(B-V)_values}. The estimated values of $E(B-V)$ based on the observed SN properties all point to substantial attenuation ($\gtrsim 0.5$ mag). The Balmer decrement also yields a large value of $0.449 \pm 0.148$ mag. The discrepancy between our Na I D estimate ($E(B-V) \sim 0.18$ mag) and the color-based estimates ($E(B-V) \sim 0.6–0.7$ mag) is consistent with broader trends observed in SESNe populations.

\subsection{Pseudo Bolometric Light Curve}
\label{sec:pseudo-bolometric}

We construct a pseudo-bolometric light curve from all the bands with detections using the \textsc{SuperBol} code \citep{2018RNAAS...2..230N}. The routine synthesizes an SED from multi-color photometry, corrected for the MW attenuation and the host attenuation. We adopt the ATLAS $o$-band as a reference for extrapolating flux in other bands and interpolate between the observed filters to build a time-series SED. The fluxes at early and late epochs are extrapolated by assuming a constant color relative to the ATLAS-o reference band. 

We compute the pseudo-bolometric light curve and perform a blackbody fit, assuming $E(B-V)$ = 0 mag and adopting representative $E(B-V)$ values derived from the Na I D EW, the color evolution method ($0–20$ d), and the spectroscopic continuum dereddening from Table \ref{tab:E(B-V)_values}. We exclude the color-template (+10~d) estimate, as its value is intermediate between the latter two methods. The blackbody temperature and radius at the epoch of the $g$-band peak, together with the peak luminosity, are reported in Table \ref{tab:Psuedo_bol_values}.

\begin{figure}[ht!]
\centering
\includegraphics[width=0.6\linewidth]{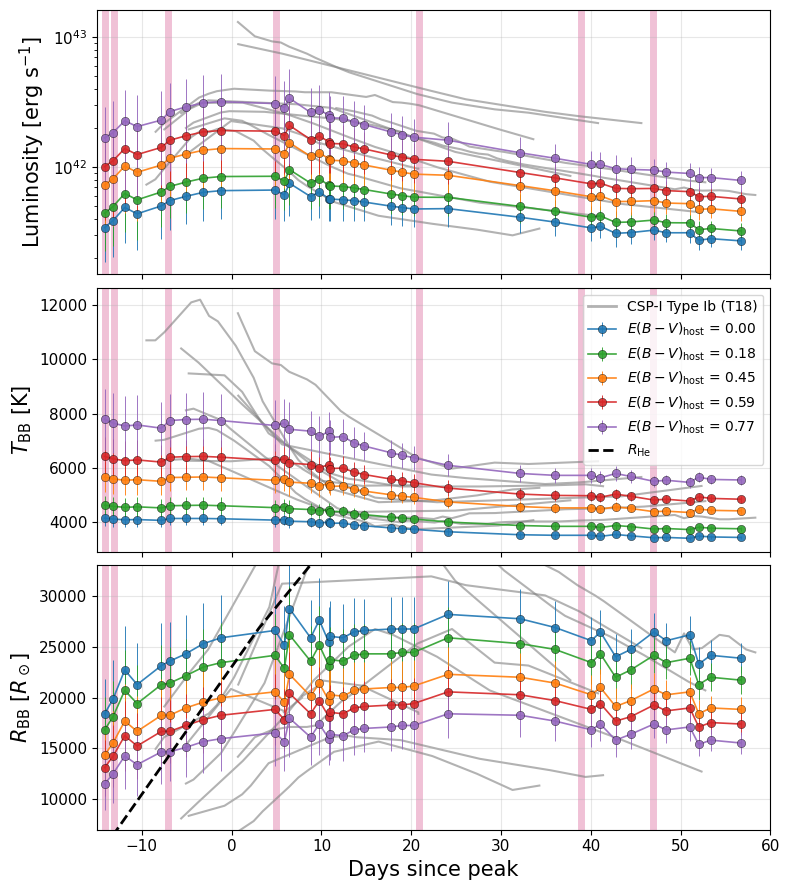}
\caption{Pseudo-bolometric light curve and blackbody evolution of SN~2024vjc under different host-attenuation assumptions, derived with \textsc{SuperBol} \citep{2018RNAAS...2..230N}. From top to bottom, the panels show luminosity, blackbody temperature, and blackbody radius as a function of phase relative to the LC peak. Different colors correspond to different assumed values of $E(B-V)_{\mathrm{host}}$. Gray lines in the background show the CSP-I Type~Ib sample from T18, included for comparison. The dashed black line in the bottom panel indicates the radius of the He line-forming region (R$_{\mathrm{He}}$), computed as v$_{\mathrm{He}}$. Vertical pink lines mark the epochs of spectroscopic observations.}  
\label{fig:bol_lc}
\end{figure}

Figure~\ref{fig:bol_lc} presents the resulting pseudo-bolometric light curve and blackbody parameters under the five host-attenuation assumptions considered. Without host attenuation, SN~2024vjc reaches a peak luminosity of $\sim 7.5\times10^{41}$~erg~s$^{-1}$, with a blackbody temperature of $\sim 4100$~K at $g$-band peak that gradually decreases afterward, and a blackbody radius of $\sim 26000~R_{\odot}$ at $g$-band peak. As the assumed host attenuation increases, the inferred peak luminosity rises by a factor of $\sim 4.5$, the blackbody temperature at $g$-band peak shifts to higher values (reaching $\sim 6400$~K for $E(B-V)_{\mathrm{host}} = 0.59$~mag, and $\sim 7700$~K for $E(B-V)_{\mathrm{host}} = 0.77$~mag), and the inferred radius decreases correspondingly via the Stefan-Boltzmann relation. The extinction-corrected luminosity and temperature are consistent with those of typical stripped-envelope supernovae, as illustrated by the CSP-I Type~Ib comparison from T18 shown in the background of Figure~\ref{fig:bol_lc} and results in the derived blackbody temperature ($\sim$ 4100 K) significantly lower than temperatures typically observed in SESNe near maximum light. On the other hand, the no-extinction assumption places SN~2024vjc well below the T18 distribution in both luminosity and blackbody temperature; as shown in the figure, even the coolest objects in the T18 sample reach $\sim$6000~K near maximum light, approximately 1500~K hotter than the inferred temperature of SN~2024vjc under the assumption of no host attenuation.


\begin{deluxetable}{lccr}
\tablecaption{SN 2024vjc Pseudo bolometric peak parameters for different values of $E(B-V)$ \label{tab:Psuedo_bol_values}}
\tablehead{
\colhead{$E(B-V)$ [mag]} & \colhead{Temperature at $g$-band peak [K]} & \colhead{Radius at $g$-band peak [R$_{\odot}$]} & \colhead{Peak Luminosity [erg s$^{-1}$]}}
\startdata
0 & 4120  &  25\,873  & 7.53 $\times$ 10$^{41}$ \\
0.18 & 4600  &  23\,430 & 9.58 $\times$ 10$^{41}$ \\
0.45 & 5630 &  19\,980 & 1.54 $\times$ 10$^{42}$ \\
0.59  & 6390 &  18\,255 & 2.09 $\times$ 10$^{42}$ \\
0.77 & 7740 &  15\,955 & 3.38 $\times$ 10$^{42}$ \\
\enddata
\tablecomments{Values obtained using \textsc{SuperBol} }
\end{deluxetable}



In addition, the evolution of the photospheric radius can provide another constraint: the photosphere must form below the He line-forming region. Since the blackbody radius derived with \textsc{SuperBol} provides an approximation to the photospheric radius ($R_{\rm BB} \approx R_{\rm phot}$), it must also remain below the radius of the He line-forming region. To investigate this constraint, the bottom panel of Figure~\ref{fig:bol_lc} shows the radius of the He line-forming region assuming homologous expansion with representative He velocities, providing a comparison scale for the inferred photospheric radius. The constraint is not satisfied for the no-host-attenuation case, where $R_{\rm BB}$ exceeds the line-forming region in the pre-maximum phase. For higher attenuation values, $R_{\rm BB}$ decreases due to the rising temperature, and for $E(B-V)_{\mathrm{host}} \gtrsim 0.45$~mag the inferred radius stays below the upper limit set by $R_{\rm He}$ at all observed epochs. This independently supports the conclusion that SN~2024vjc suffers from substantial host-galaxy attenuation.


\section{Discussion} \label{sec:discussion}



\subsection{Substantial host attenuation without a trace of Na ID absorption}


The Na~I~D diagnostic was originally calibrated for SNe~Ia due to their uniform observational properties. While the method has several known limitations, it has been regarded as a useful indicator -- especially, the absence of absorption is often interpreted as evidence for negligible host-galaxy attenuation \citep{2013ApJ...779...38P}.

Because CCSNe occur in more complex environments than SNe~Ia, this assumption requires independent testing. In the present work, we find that SN~Ib~2024vjc serves as a cautionary case for this diagnostic. The Na~\textsc{i}~D features are only marginally detected, and the standard calibration of \citet{2012MNRAS.426.1465P} yields a low color excess of $E(B-V)_{\rm host} \sim 0.18$~mag. The alternative \citet{2003fthp.conf..200T} and not as widely used as \citet{2012MNRAS.426.1465P} Na~\textsc{i}~D calibration yields a higher value ($E(B-V)_{\rm host} \approx 0.44$~mag). All SN-based indicators (color-template matching, spectroscopic continuum modeling) and the Balmer decrement consistently point toward significant attenuation in the range $\sim 0.45$--$0.77$~mag. 

Rather than identifying a single, definitive value for $E(B-V)_{\mathrm{host}}$, our results highlight the inherent systematic uncertainties across different methodologies. While the exact value is difficult to constrain precisely, the consensus across multiple diagnostics places SN~2024vjc in a high-attenuation regime ($E(B-V)_{\mathrm{host}} \approx 0.5$ $-$ $0.8$~mag). We therefore recommend that extinction estimates for CCSNe not rely solely on Na~I~D, but instead be treated as a range derived from complementary analyses, such as color evolution, excitation temperatures, and environmental context.

A close look at physics-motivated (assumption-free) consistency checks is essential when deriving these properties. In this study, we also cross-checked our results in this aspect, including the predicted photospheric temperature compared to the required excitation temperature, the predicted photospheric velocity compared to the location of the line-forming region.
Together with the above mentioned various diagnostics, the collective agreement on a high-extinction scenario reinforces the physical reality of the reddening.


\subsection{Discussion on strong reddening and weak Na I D absorption} \label{sec:Discussion_environment}

A natural question raised by the present result is why SN 2024vjc exhibits strong reddening while showing only weak Na I D absorption. \citet{2012MNRAS.426.1465P} derived a theoretical relation linking the equivalent width of Na I D to $E(B-V)$, given by their Equation 6:
\begin{equation} \label{eq:EW_EBV_relation}
\mathrm{EW(Na I D)} \propto \kappa_{0} (1 - f_{\mathrm{ion}}) f_{\mathrm{Na}} f_{\mathrm{g2d}} F(R_{V}, E(B-V)) \ ,
\end{equation}
where $\kappa_{0}$ is the absorption coefficient, $f_{\mathrm{ion}}$ is the sodium ionization fraction, $f_{\mathrm{Na}}$ is the sodium abundance relative to the gas, $f_{\mathrm{g2d}}$ is the gas-to-dust ratio, and $F(R_{V}, E(B-V))$ is an arbitrary function encapsulating the attenuation. As noted by \citet{2012MNRAS.426.1465P}, despite the large number of variables and their known diversity even within the MW, it is remarkable that the correlation between EW(Na I D) and $E(B-V)$ survives at all.

By examining the individual terms in Equation \ref{eq:EW_EBV_relation}, plausible explanations for the discrepancy observed in SN 2024vjc emerge. In our photometric analysis we adopt $R_V = 3.1$, consistent with the MW–like ratio of total to selective attenuation. However, this value may not be representative for external galaxies \citep[e.g.,][]{2015A&A...577A..53P}. To check this, in our spectroscopic de-reddening analysis, we allowed both $A_V^{*}$ and $R_V^{*}$ to vary simultaneously, and found that a high value of $E(B-V)$ value is required for a range of $R_{V}^{*}$. This suggests that the strong reddening inferred for SN 2024vjc is not primarily driven by an incorrect assumption about the attenuation encapsulated in $F(R_{V}$, $E(B-V)$). Instead, the discrepancy could be explained by variations in the other terms of Equation~\ref{eq:EW_EBV_relation}. A combination of sodium poor gas (low $f_{\mathrm{Na}}$) and dust-rich material, corresponding to a reduced gas-to-dust ratio (low $f_{\mathrm{g2d}}$), can weaken the Na~I~D absorption while leaving the dust attenuation largely unaffected. A more likely solution would indeed be the ionization effect on sodium -- a high sodium ionization fraction ($f_{\mathrm{ion}}$) would suppress Na~I~D absorption even in the presence of substantial dust. 
We note that such conditions are found in the warm interstellar medium, where sodium exists primarily as Na II despite the presence of dust. Such environments are characterized by low $N(\mathrm{Na~I})/N(\mathrm{Ca~II})$ ratios, as demonstrated by \citet{2010A&A...510A..54W}.

The opposite behavior has also been observed in SNe Ia. \citet{2013ApJ...779...38P} showed that approximately 25$\%$ of SNe Ia exhibit anomalously strong Na I D absorption relative to their measured dust attenuation, deviating significantly from the standard MW relation. This excess Na I D absorption may arise from different physical conditions, including higher metallicity environments and the difference in circumstellar material associated with the progenitor system. In contrast, CCSNe such as SN 2024vjc occur in younger, more actively star forming regions \citep{2012MNRAS.424.1372A}, where enhanced dust content, stronger radiation fields, and ionized gas can suppress Na I D absorption while still reddening the object. Together, these examples highlight the complexity of the relation between the Na~I~D absorption and the attenuation, and demonstrate that Na~I~D absorption alone cannot be considered a reliable proxy for dust attenuation across different SN types and environments.

This conclusion is further reinforced by \citet{2026A&A...707A.272G}, who analyzed narrow interstellar absorption lines in over 1800 low-redshift SNe, finding that SESNe show systematically stronger Na~{\sc i}~D EW than SNe~II despite similar star-forming environments — a result they attribute to local material near the explosion site rather than the integrated host galaxy ISM. SN~2024vjc may represent the complementary case: the ionizing radiation field of the H~{\sc ii} region suppresses neutral sodium, producing anomalously weak absorption despite substantial dust. Taken together, these results demonstrate that Na~{\sc i}~D in SESNe can deviate from the standard calibration in both directions, reinforcing caution against its use as a reliable attenuation proxy for this SN class.

\subsection{Evidence for a Hot Star-Forming Region at the SN Site}
\label{sec:Disc_HII}

The ionization scenario proposed in Section~\ref{sec:Discussion_environment} predicts the presence of hot, young stars at the SN site. To test this, we examine the late-time GMOS spectrum obtained at $+123$\,d. At this epoch, the supernova continuum has faded sufficiently to unveil the underlying stellar population and any associated emission from the host environment.

\begin{figure}[ht!]
\centering
\includegraphics[width=0.75\linewidth]{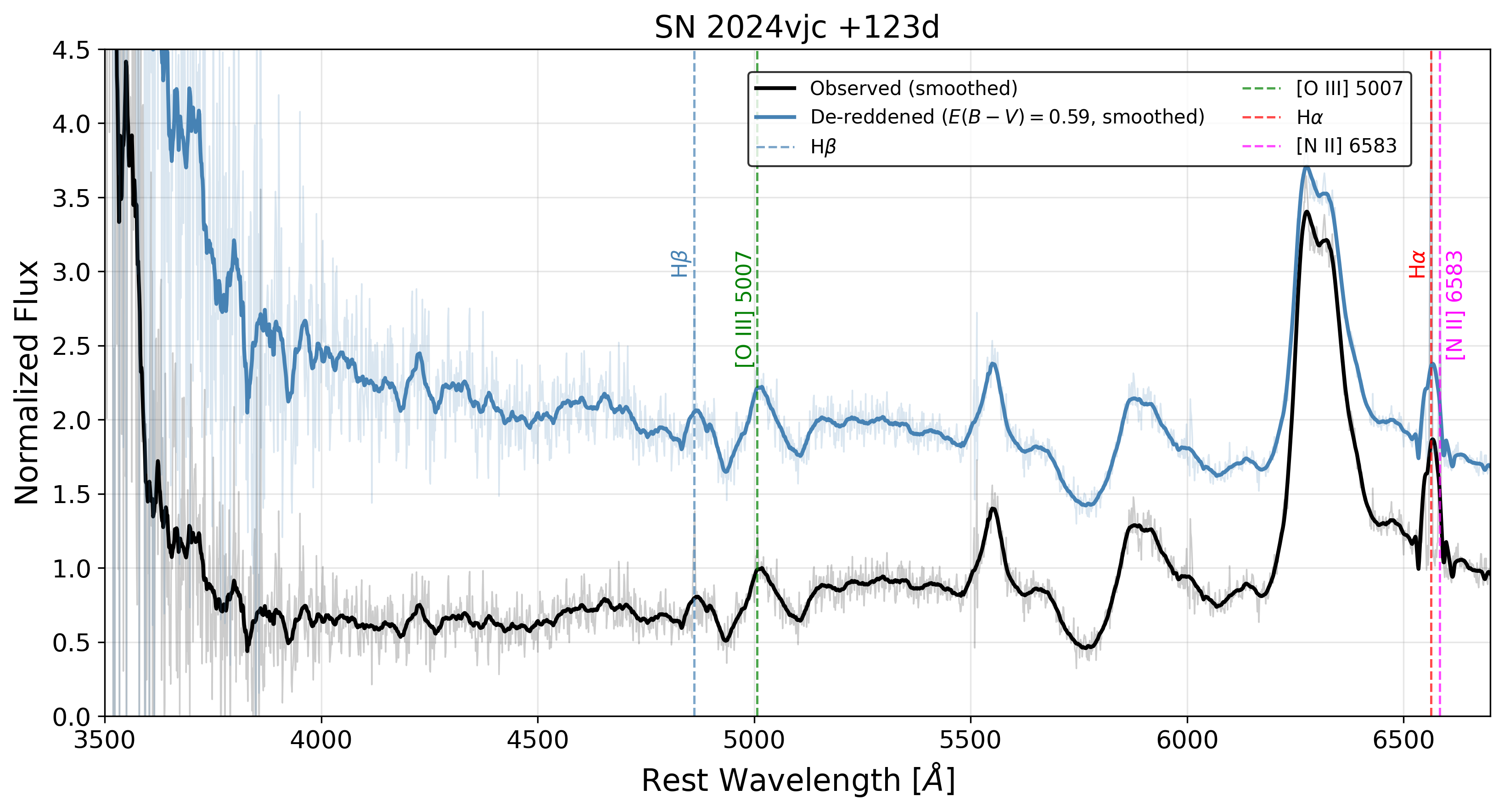}
\caption{The $+123$\,d observed spectrum is shown in gray (raw) and black (smoothed). The de-reddened spectrum, corrected assuming $E(B-V) = 0.59$ mag with a $R_{V} = 3.1$ CCM89 extinction law, is shown in light blue (raw) and blue (smoothed); both spectra are normalized to unity in the 5400--5600\,\AA\ region and offset. Vertical dashed lines mark the rest-frame wavelengths of H$\beta$\,$\lambda$4861 (blue dashed), [\ion{O}{3}]\,$\lambda$5007 (green dashed; shown for reference only -- see Section~\ref{sec:Disc_HII} for discussion of why the observed feature near this wavelength is inconsistent with an \ion{H}{2} region origin), H$\alpha$\,$\lambda$6563 (red dashed), and [\ion{N}{2}]\,$\lambda$6583 (magenta dashed). We note that the other lines are resolved in the original, non-smoothed spectra (Fig.~\ref{fig:emission_fits}).}
\label{fig:blue_continuum_hii}
\end{figure}

Figure~\ref{fig:blue_continuum_hii} shows the $+123$\,d GMOS-S spectrum of SN~2024vjc, both observed and de-reddened assuming $\mathrm{E}(B-V) = 0.59$\,mag, alongside the rest-frame positions of key \ion{H}{2} region diagnostic lines. The de-reddened spectrum reveals a blue continuum excess below $\sim$4500\,\AA.

\begin{figure}[ht!]
\centering
\includegraphics[width=0.95\linewidth]{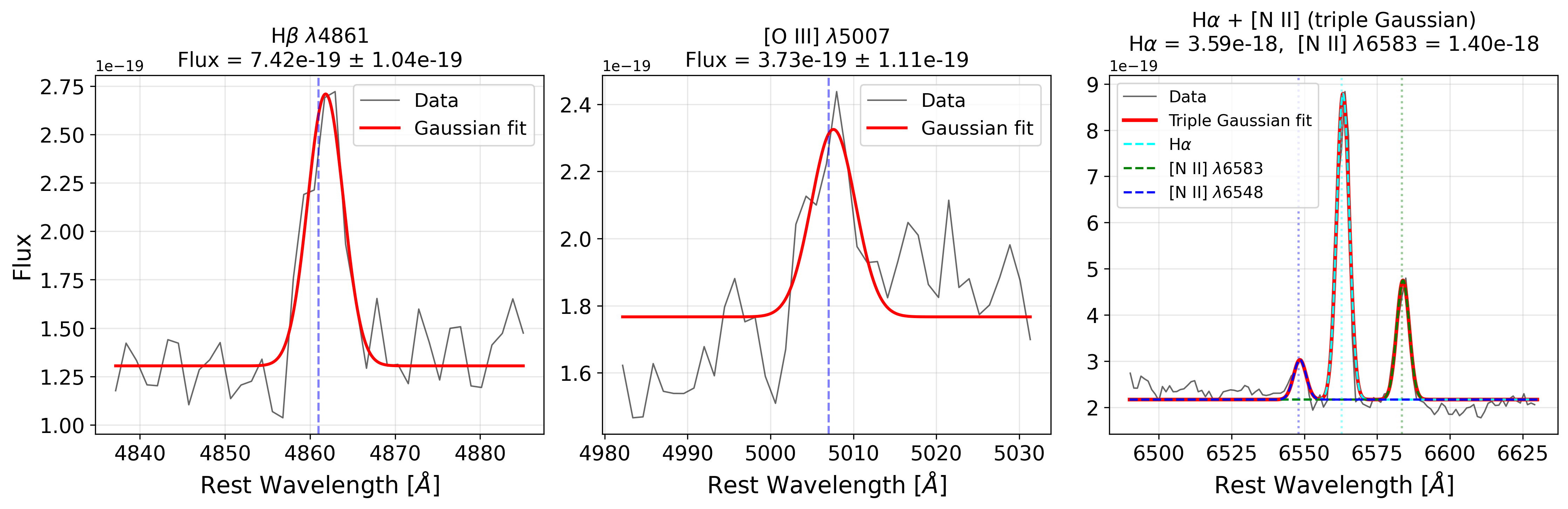}
\caption{Emission line fits to the $+123$\,d GMOS-S spectrum of SN~2024vjc. \textit{Left:} H$\beta$\,$\lambda$4861 fitted with a single Gaussian (red curve). \textit{Center:} The feature near 5007\,\AA\ fitted with a single Gaussian (red curve); we caution that its width appears inconsistent with an H~II region [\ion{O}{3}]\,$\lambda$5007 identification. \textit{Right:} The H$\alpha$ + [\ion{N}{2}] complex fitted simultaneously with a physically constrained triple Gaussian model (red curve). Individual components are shown for H$\alpha$ (cyan dashed), [\ion{N}{2}]\,$\lambda$6583 (green dashed), and [\ion{N}{2}]\,$\lambda$6548 (blue dashed). The gray curve shows the observed spectrum in all panels. The derived metallicity diagnostics and H$\alpha$ equivalent width are reported in Table~\ref{tab:metallicity}.}
\label{fig:emission_fits}
\end{figure}

The slope close to $\sim$3500\,\AA\ may be partially affected by the blue edge of the GMOS wavelength coverage, so we interpret this feature with caution. Nonetheless, the presence of excess blue flux supports the existence of a young, hot stellar population in the vicinity of the SN site. Narrow emission lines are clearly detected at H$\alpha$\,$\lambda$6563, [\ion{N}{2}]\,$\lambda\lambda$6548,\,6583, and a feature consistent with H$\beta$\,$\lambda$4861 is also present (Figure~\ref{fig:emission_fits}). A feature near $\sim$5007\,\AA\ is also present, but its profile appears broader than the detected H~II region emission lines (H$\alpha$, [\ion{N}{2}], H$\beta$) and is therefore inconsistent with an [\ion{O}{3}]\,$\lambda$5007 origin from an H~II region; we discuss this further below. The detected narrow H$\alpha$, [\ion{N}{2}], and H$\beta$ emission lines are consistent with an underlying \ion{H}{2} region in the projected vicinity of SN~2024vjc.

To characterize the properties of the star-forming environment, we measure the fluxes of the detected emission lines using Gaussian fits (Figure~\ref{fig:emission_fits}). The metallicity is estimated using the N2 strong-line diagnostic of \citet{2013AaA...559A.114M}, which depends only on the unambiguously detected H$\alpha$ and [\ion{N}{2}] lines. We obtain $12+\log(\mathrm{O/H}) \approx 8.55$ from the N2 index [$\log([\mathrm{N\,II}]\,\lambda6583/\mathrm{H}\alpha) = -0.41$], indicating a moderately sub-solar metallicity of $\sim$$0.7\,Z_\odot$ ($12+\log(\mathrm{O/H})_\odot = 8.69$). We do not adopt the O3N2 diagnostic, as the feature near 5007\,\AA\ in our spectrum appears broader than the other H~II region emission lines (H$\alpha$, [\ion{N}{2}], H$\beta$) and is inconsistent with the velocity dispersion expected for ionized gas in an H~II region. This feature may instead reflect residual SN ejecta emission or possible contamination by He~I~$\lambda$5016, and we therefore consider the O3N2-based metallicity (reported in Table~\ref{tab:metallicity} for completeness) to be unreliable.

We measure an H$\alpha$ EW of $\mathrm{EW(H\alpha)} = 16.6 \pm 0.8$\,\AA. This value should be treated as a lower limit, as the SN continuum still contributes and dilutes the true \ion{H}{2} region equivalent width \citep[see e.g.][]{2018A&A...613A..35K}. The true EW(H$\alpha$) of the underlying \ion{H}{2} region will only be measurable once the SN has fully faded. Compared with the environment samples of \citet{2018A&A...613A..35K} and \citet{2023A&A...677A..28P}, the measured EW(H$\alpha$) is consistent with the typical range for SESNe, while the metallicity places SN~2024vjc at the high-metallicity end of both samples, possibly above the bulk of the \citet{2023A&A...677A..28P} Type~Ib distribution.

The Balmer decrement measured from these emission lines ($E(B-V)_{\rm Balmer} = 0.449 \pm 0.148$\,mag; Section~\ref{sec:balmer}) provides an additional perspective on the dust distribution around SN~2024vjc. This value is intermediate between the Na\,\textsc{i}\,D estimate ($E(B-V) = 0.18$\,mag) and the SN-based methods ($E(B-V) = 0.59$--$0.77$\,mag). One possible interpretation is that this ordering reflects the different physical scales each method probes: the Na\,\textsc{i}\,D absorption traces neutral gas on giant molecular cloud scales; the Balmer decrement probes dust on the scale of the stellar cluster hosting the progenitor; while the SN light-curve and spectral methods are sensitive to dust on the scale of the immediate circumstellar environment. If this picture is correct, the trend would be suggestive of a centrally concentrated dust distribution around the explosion site, with the bulk of the attenuation residing on smaller scales closer to the progenitor. 

Taken together, the detection of narrow Balmer and forbidden emission lines, the blue continuum excess in the de-reddened spectrum, and the moderately sub-solar local metallicity ($\sim$$0.7\,Z_\odot$) all support the presence of an active \ion{H}{2} region in the projected vicinity of SN~2024vjc, typical of environments seen for SESNe. We note, however, that without three-dimensional spatial information we cannot confirm that the SN progenitor was physically embedded within this \ion{H}{2} region, as a chance projected alignment cannot be excluded. Nevertheless, this is consistent with our proposed scenario in which a strong local ionizing radiation field from hot young stars photoionizes the sodium along the line of sight, suppressing Na\,\textsc{i}\,D absorption despite the presence.

\begin{deluxetable}{lcc}
\tablecaption{Emission line measurements and metallicity diagnostics 
for SN~2024vjc at $+123$\,d from the GMOS-S spectrum. 
\label{tab:metallicity}}
\tablehead{
\colhead{Quantity} & \colhead{Value} & \colhead{Unit}
}
\startdata
\multicolumn{3}{c}{\textit{Line Fluxes}} \\
\hline
H$\beta$ $\lambda$4861              & $(7.42 \pm 1.04) \times 10^{-19}$ & erg s$^{-1}$ cm$^{-2}$ \\
{[\ion{O}{3}]} $\lambda$5007$^{\dagger}$     & $(3.73 \pm 1.11) \times 10^{-19}$ & erg s$^{-1}$ cm$^{-2}$ \\
H$\alpha$ $\lambda$6563             & $(3.59 \pm 0.37) \times 10^{-18}$ & erg s$^{-1}$ cm$^{-2}$ \\
{[\ion{N}{2}]} $\lambda$6583        & $(1.40 \pm 0.15) \times 10^{-18}$ & erg s$^{-1}$ cm$^{-2}$ \\
\hline
\multicolumn{3}{c}{\textit{Metallicity Diagnostics}} \\
\hline
N2 index                            & $-0.409 \pm 0.039$             & \nodata \\
$12+\log(\mathrm{O/H})$ [N2]        & $8.5539 \pm 0.0181$             & dex     \\
O3N2 index$^{\dagger}$                     & $0.110 \pm 0.148$     & \nodata \\
$12+\log(\mathrm{O/H})$ [O3N2]$^{\dagger}$  & $8.5094 \pm 0.0317$   & dex     \\
\hline
\multicolumn{3}{c}{\textit{H$\alpha$ Equivalent Width}} \\
\hline
EW(H$\alpha$)$^{*}$                 & $16.6 \pm 0.8$        & \AA     \\
\enddata
\tablecomments{Line fluxes are not corrected for reddening. Metallicity calibrations follow \citet{2013AaA...559A.114M}. Solar metallicity is $12+\log(\mathrm{O/H})_\odot = 8.69$. $^{*}$EW(H$\alpha$) is a lower limit as the SN continuum still contributes at $+123$\,d, diluting the true \ion{H}{2} region equivalent width. $^{\dagger}$The [\ion{O}{3}] $\lambda$5007 flux and O3N2-based values are reported for completeness but are unreliable due to the ambiguous identification of the 5007\,\AA\ feature. The adopted metallicity for SN~2024vjc is the N2 value, $12+\log(\mathrm{O/H}) = 8.55$.}
\end{deluxetable}

\subsection{Comparison to SESN sample studies}

In this section, we place our results in the context of previous studies of stripped-envelope supernovae. Using the color evolution method of \citet{2011ApJ...741...97D}, \citet{2022MNRAS.512.3195Z} derived host-galaxy attenuation estimates for a sample of 70 SESNe from the Lick Observatory Supernova Search, finding a mean $E(B-V)$ = 0.32 $\pm$ 0.19 mag. This value provides a useful benchmark for photometrically inferred attenuation estimates in SESNe.

\begin{figure}[ht!]
\centering
\includegraphics[width=0.75\linewidth]{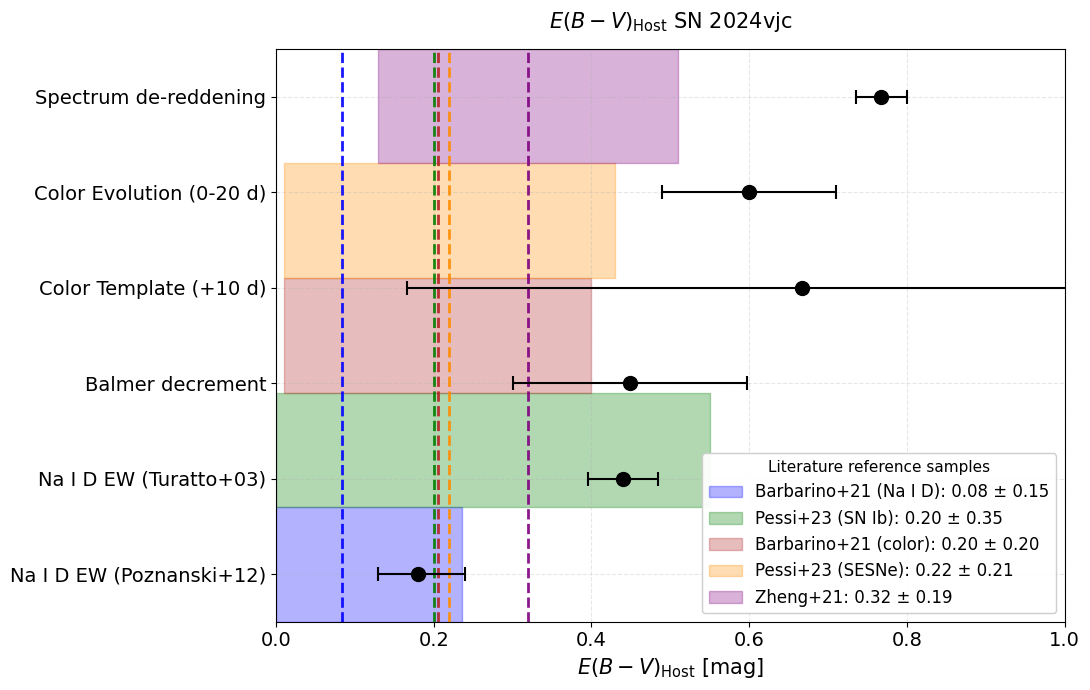}
\caption{Summary of the host-galaxy color excess estimates for SN~2024vjc obtained using different methods. Black points indicate the inferred $E(B-V)_{\mathrm{host}}$ values, with horizontal error bars representing the associated uncertainties (see Table~\ref{tab:E(B-V)_values}). Colored shaded regions and dashed vertical lines indicate literature reference values: the blue and red bands show the mean values from the Type~Ic sample of \citet{2021A&A...651A..81B} estimated from Na\,\textsc{i}\,D absorption \citep[][ calibration]{2012MNRAS.426.1465P} and from photometric $g-r$ colors, respectively; the green and orange bands show the median $E(B-V)$ values and dispersions for SNe~Ib and SESNe from \citet{2023A&A...677A..28P}; and the purple band corresponds to the mean $E(B-V)$ and dispersion for SESNe reported by \citet{2022MNRAS.512.3195Z}  
}
\label{fig:Literature_attenuation}
\end{figure}

An independent perspective is provided by \citet{2023A&A...677A..28P}, who studied the environments of 112 core-collapse SNe using VLT/MUSE observations. Host-galaxy attenuation was estimated from the Balmer decrement measured in nearby H II regions, a method that probes the SN environment directly and does not rely on the SN’s photometric or spectroscopic properties. For SNe Ib, they report a median $E(B-V)$ = 0.20 $\pm$ 0.35 mag and a mean $E(B-V)$ = 0.26 $\pm$ 0.28 mag, while SESNe as a class show median and mean values of 0.22 $\pm$ 0.21 mag and 0.30 $\pm$ 0.16 mag, respectively.

A direct illustration of the systematic offset between Na\,\textsc{i}\,D and color-based diagnostics in SESNe is provided by the Type~Ic sample of \citet{2021A&A...651A..81B}, who derive host-galaxy attenuation for 44 PTF/iPTF SNe~Ic using both methods. The Na\,\textsc{i}\,D-based estimates yield a mean $E(B-V) = 0.08 \pm 0.15$~mag, while the $g-r$ color-based estimates give a mean of $0.20 \pm 0.20$~mag, indicating that color-based methods systematically infer larger attenuation than Na\,\textsc{i}\,D in this class. SN~2024vjc represents an extreme example of this same trend.

The values of $E(B-V)$ inferred for SN~2024vjc are compared to these statistical samples in Figure~\ref{fig:Literature_attenuation}. Notably, the Na\,\textsc{i}\,D estimate based on the alternative \citet{2003fthp.conf..200T} calibration (Section~\ref{sec:Na_Id_attenuation}) yields $E(B-V)=0.444 \pm 0.044$~mag, substantially higher than the \citet{2012MNRAS.426.1465P} value and in good agreement with the Balmer decrement result. The high attenuation derived from the SN-based and Balmer decrement diagnostics, $E(B-V)=0.45-0.77$~mag, places SN~2024vjc well above the typical range inferred for SESNe by all reference samples. If only the Na\,\textsc{i}\,D EW were considered using the standard calibration, SN~2024vjc would appear consistent with the median values reported by previous works.

A single object cannot, however, distinguish between two interpretations of this finding. SN~2024vjc may be a genuine outlier -- an unusually heavily-attenuated object in an unusual environment -- in which case the Na\,\textsc{i}\,D diagnostic remains reliable for the bulk of the SESN population and our result is primarily a cautionary case. Alternatively, SN~2024vjc may represent a hidden subpopulation of highly-attenuated SESNe that have so far escaped attention precisely because their Na\,\textsc{i}\,D EWs do not flag them as reddened, leading to a systematic underestimation of the high-attenuation tail of the SESN distribution. The diverse environmental conditions in which SNe~Ib (and SESNe more generally) explode make either scenario plausible. Distinguishing between them will require a systematic search for similar objects in larger, homogeneously analyzed samples. The present work demonstrates that such objects exist; quantifying their frequency in the SESN population remains an open question, and establishing robust attenuation indicators beyond the Na\,\textsc{i}\,D diagnostic is essential to address it.

\section{Conclusions} \label{sec:conclusion}


We present a multi-band photometric and spectroscopic study of the Type~Ib supernova SN~2024vjc, covering the evolution from shortly after explosion to $\sim$130~days. While its rise time, decline rate, and spectral morphology are broadly consistent with a standard Type~Ib SN, SN~2024vjc would be an extreme outlier—among the faintest and reddest objects in the SESN population—if no host-galaxy attenuation were assumed.

A central result of this work is the significant discrepancy between different host-galaxy attenuation diagnostics. While the Na\,\textsc{i}\,D~EW method using the \citet{2012MNRAS.426.1465P} calibration yields a modest value of $E(B-V)_{\mathrm{host}} \sim 0.18$~mag, all SN-based diagnostics (color templates, color evolution, and spectroscopic continuum comparisons) consistently indicate substantially higher attenuation in the range of $E(B-V)_{\mathrm{host}} \sim 0.6$--$0.8$~mag. The alternative Na\,\textsc{i}\,D calibration of \citet{2003fthp.conf..200T} yields $E(B-V)_{\mathrm{host}} \approx 0.44$~mag, in good agreement with the Balmer decrement estimate. We further note that the Na\,\textsc{i}\,D measurement is complicated by an anomalous doublet ratio EW(D2)/EW(D1)~$\approx 1$, which points to either saturation of the lines or contamination from the broad He\,\textsc{i}\,$\lambda 5876$ line of the SN, both of which limit the reliability of the standard EW--$E(B-V)$ calibration in this case.


Multiple independent lines of evidence favor the high attenuation scenario:
\begin{enumerate}
\item 
A normal SN Ib with the high attenuation provides a consistent picture for essentially all the observed features except only for the Na ID absorption, providing the most straightforward interpretation. 
\item The spectral continuum shape is better reproduced by applying attenuation than by invoking an intrinsically cooler single blackbody temperature.
\item The black body radius (which is $\sim$ photosphere radius) inferred from the blackbody fit is consistent with the constraint imposed by the He\,I line velocities only under the high-attenuation assumption. 
\item The excitation temperature required by the \textsc{SYN++} spectral modeling is more consistent with the blackbody temperature evolution obtained when host attenuation is included.
\end{enumerate}

Late-time spectroscopy at $+123$\,d reveals narrow H$\alpha$, [\ion{N}{2}], and H$\beta$ emission lines, together with a blue continuum excess in the de-reddened spectrum, consistent with an active \ion{H}{2} region in the vicinity of the explosion site. This environment provides a hint about a possible cause of the weak Na~{\sc i}~D; the strong ionizing radiation field from hot young stars in this region may plausibly contribute to photoionizing the neutral sodium, suppressing the Na~{\sc i}~D feature despite the presence of substantial dust. 

The Balmer decrement measured from narrow emission lines in late-time spectra ($E(B-V)_{\rm Balmer} = 0.449 \pm 0.148$~mag; Section~\ref{sec:balmer}) provides a crucial intermediate value. One possible interpretation is that the ordering of these estimates $-$ increasing from the Na~I~D diagnostic, through the Balmer decrement, to the SN-based methods $-$ could reflect the different physical scales probed by each diagnostic. In such a scenario, Na~I~D absorption would trace neutral gas on the scale of the diffuse interstellar medium or giant molecular clouds; the Balmer decrement would probe dust associated with the stellar cluster or \ion{H}{2} region hosting the progenitor; and the SN-based methods would be sensitive to the immediate circumstellar environment. If this picture holds, it would be suggestive of a centrally concentrated dust distribution around the explosion site, with a significant fraction of the total attenuation residing on small scales near the progenitor.

We note that some individual properties of SN~2024vjc depart from the average behavior of the SN~Ib population, which warrants some caution in extrapolating our conclusions. The $r-i$ color evolution is unusually flat between $-10$ and $+20$~d relative to peak, and the He~{\sc i} $\lambda$5876 velocities lie toward the high end of the SESN distribution (Section~\ref{sec:Spectra}). The small excess in the $i$- and $z$-band light curves near MJD $\sim$60\,625 coincides with the emergence of nebular Ca~{\sc ii} features (Section~\ref{sec:light_curve}); a qualitatively similar excursion appears in a preliminary comparison with several literature SESNe, though we do not attempt to quantify how common this feature is, given the general scarcity of well-sampled $i$-band photometry through this phase in the literature. While the combination of independent SN-based diagnostics presented here provides a consistent and compelling case for substantial host-galaxy attenuation, because SN~2024vjc has somewhat unusual properties compared with typical SESNe, some of our extinction estimates could be affected by systematic biases. To confirm that our conclusions are generally valid, we need a larger sample of similarly well-observed, heavily dust-obscured SESNe.

Nonetheless, SN~2024vjc represents a clear counterexample to the commonly adopted assumption that weak or absent Na~{\sc i}~D absorption implies negligible host-galaxy attenuation. The ionization scenario as discussed above may represent a general caveat for Na~{\sc i}~D-based attenuation estimates in SESNe occurring in actively star-forming environments. Our results reinforce the need for caution when using Na~I~D as an attenuation proxy and highlight the importance of employing multiple, independent diagnostics when estimating attenuation in CCSNe. Overall, this study underscores the complexity of dust attenuation in SN environments and demonstrates that heavily attenuated SESNe can remain hidden among apparently non attenuated events if Na~I~D–based diagnostics are used in isolation.

\section*{Data Availability}

The photometric data underlying Figure~\ref{fig:sn24vjc_lc} are available as Data behind the Figure (DbF), including a machine-readable table and accompanying README describing its contents. The spectroscopic observations of SN~2024vjc presented in this work will be made available via the Weizmann Interactive Supernova data REPository \citep[WISeREP;][]{2012PASP..124..668Y}\footnote{\url{https://www.wiserep.org}}.

\begin{acknowledgments}
We thank M. Fraser for insightful comments on the Na\,\textsc{i}\,D analysis presented in this work. K.M. acknowledges support from JSPS KAKENHI grant (JP24KK0070, JP24H01810, JP24K00682, and JP23H04894). The work is partly supported by the JSPS Open Partnership Bilateral Joint Research Projects; between Japan and Finland (JPJSBP120229923) and between Japan and Chile (JPJSBP120259901). 
A.R., G.V. acknowledge financial support from the SOXS project (PI S. Campana). 
A.R., A.P., G.V. acknowledge financial support from the PRIN-INAF 2022 "Shedding light on the nature of gap transients: from the observations to the models".
H.K. was funded by the Research Council of Finland projects 324504, 328898, and 353019. 
The observations by Gemini-S (S24B-041, GS-2024B-Q-101, GS-2024B-Q-102; PI: K. Maeda) were carried out within the framework of the Subaru-Keck/Subaru-Gemini time exchange program, which is operated by the National Astronomical Observatory of Japan. We are honored and grateful for the opportunity of observing the Universe from Maunakea, which has cultural, historical, and natural significance in Hawaii. This article is also based on observations obtained from the La Silla Observatory with the REM telescope, under the program REM AOT47-37 (ID 49337, PI: G. Valerin). This work has made use of data from the Asteroid Terrestrial-impact Last Alert System (ATLAS) project. The Asteroid Terrestrial-impact Last Alert System (ATLAS) project is primarily funded to search for near earth asteroids through NASA grants NN12AR55G, 80NSSC18K0284, and 80NSSC18K1575; byproducts of the NEO search include images and catalogs from the survey area. This work was partially funded by Kepler/K2 grant J1944/80NSSC19K0112 and HST GO-15889, and STFC grants ST/T000198/1 and ST/S006109/1. The ATLAS science products have been made possible through the contributions of the University of Hawaii Institute for Astronomy, the Queen’s University Belfast, the Space Telescope Science Institute, the South African Astronomical Observatory, and The Millennium Institute of Astrophysics (MAS), Chile. This work is based on observations collected at the European Southern Observatory (ESO) under ePESSTO+ (the advanced Public ESO Spectroscopic Survey for Transient Objects). ePESSTO+ observations were obtained under ESO program IDs 112.25JQ.011 through 112.25JQ.015 (PI: C. Inserra). Based on observations collected with the SPECULOOS and VST telescopes at Paranal Observatory, Chile, under the programme allocated by the Chilean Telescope Allocation Committee (CNTAC), no: CN2024B-52 (PI G. Pignata). Based on observations made with the Nordic Optical Telescope, owned in collaboration by the University of Turku and Aarhus University, and operated jointly by Aarhus University, the University of Turku and the University of Oslo, representing Denmark, Finland and Norway, the University of Iceland and Stockholm University at the Observatorio del Roque de los Muchachos, La Palma, Spain, of the Instituto de Astrofisica de Canarias. The NOT data were obtained under program IDs P68-505 and P70-503. The data presented here were obtained with ALFOSC, which is provided by the Instituto de Astrofisica de Andalucia (IAA) under a joint agreement with the University of Copenhagen and NOT. Observations from the NOT were obtained through the NUTS2 collaboration, which is supported in part by the Instrument Centre for Danish Astrophysics (IDA), and the Finnish Centre for Astronomy with ESO (FINCA) via Academy of Finland grant nr 306531. C.P.G. acknowledges financial support from grant RYC2024-050959-I, funded by MICIU/AEI/10.13039/501100011033 and the FSE+, as well as from projects PID2023-151307NB-I00, PIE 20215AT016, and CEX2020-001058-M, and the MaX-CSIC Excellence Award MaX4-SOMMA-ICE. T.M.R is part of the Cosmic Dawn Center (DAWN), which is funded by the Danish National Research Foundation under grant DNRF140. T.M.R acknowledges support from the Research Council of Finland project 350458. IS acknowledges financial support from the SOXS Science Consortium. T.-W.C. acknowledges financial support from the Yushan Fellow Program of the Ministry of Education, Taiwan (MOE-111-YSFMS-0008-001-P1), and from the National Science and Technology Council, Taiwan (NSTC 114-2112-M-008-021-MY3)

\end{acknowledgments}





\bibliography{sn2024vjc}{}
\bibliographystyle{aasjournalv7}



\end{document}